\documentclass[twocolumn,  tighten,  twocolappendix]{aastex631} 

\shorttitle{Environmental Dependence of Mass--size Relation of Low-mass Quiescent Galaxies}
\shortauthors{Yoon et al. (2023)}

\begin{document}

\title{Low-mass Quiescent Galaxies Are Small in Isolated Environments: Environmental Dependence of the Mass--Size Relation of Low-mass Quiescent Galaxies}

\email{yyoon@kasi.re.kr}

\author[0000-0003-0134-8968]{Yongmin Yoon}
\affiliation{Korea Astronomy and Space Science Institute (KASI), 776 Daedeokdae-ro, Yuseong-gu, Daejeon 34055, Republic of Korea}

\author[0000-0002-1710-4442]{Jae-Woo Kim}
\affiliation{Korea Astronomy and Space Science Institute (KASI), 776 Daedeokdae-ro, Yuseong-gu, Daejeon 34055, Republic of Korea}

\author[0000-0002-9434-5936]{Jongwan Ko}
\affiliation{Korea Astronomy and Space Science Institute (KASI), 776 Daedeokdae-ro, Yuseong-gu, Daejeon 34055, Republic of Korea}
\affiliation{University of Science and Technology, Gajeong-ro, Daejeon 34113, Republic of Korea}

\begin{abstract}
We study the mass--size relation of quiescent galaxies across various environments, with a particular focus on its environmental dependence at the low-mass part of $\log(M_\mathrm{star}/M_{\odot})\lesssim10.0$. Our sample consists of 13,667 quiescent galaxies with  $\log(M_\mathrm{star}/M_{\odot})\ge9.4$ and $0.01<z<0.04$ from the Sloan Digital Sky Survey. We discover that the mass--size relation of low-mass quiescent galaxies (LQGs) with $\log(M_\mathrm{star}/M_{\odot})\lesssim10.0$ depends on their environment, with LQGs in the highest-density environments exhibiting an average size $\sim70\%$ larger than those in isolated environments. Moreover, the slope of the mass--size relation for LQGs in high-density environments is significantly shallower than that of their counterparts in isolated environments. This is in contrast with high-mass quiescent galaxies with $\log(M_\mathrm{star}/M_{\odot})\gtrsim10.5$ that show a nearly identical mass--size relation across all environments. Combined with additional discoveries that the mass--size relation slopes of LQGs and star-forming galaxies are similar to each other in high-density environments, and that LQGs in higher-density environments exhibit more disk-like structures, our results support the idea that LQGs in high-density environments have evolved from star-forming galaxies through environmental effects, which are capable of causing their quenching and transformation into quiescent galaxies. With the aid of an analysis of merger rates for simulated galaxies from a cosmological galaxy formation simulation, we suggest that the steep slope and low normalization of the mass--size relation of LQGs in the lowest-density environments may originate from recent gas-rich mergers, which occur over $10$--$30$ times more frequently in the progenitors of LQGs in the lowest-density environments than in their counterparts in high-density environments at low redshifts. 
\end{abstract}
\keywords{Early-type galaxies (429) --- Galaxy environments (2029) --- Galaxy evolution (594) --- Galaxy formation (595) --- Galaxy mergers (608)  --- Galaxy structure (622) --- Quenched galaxies (2016)}

\section{Introduction}\label{sec:intro}
Stellar masses/luminosities and sizes are the most fundamental properties of galaxies, which are cumulative consequences of the series of events during the evolution and formation histories of galaxies, such as the formation of stars, mergers with other galaxies, and interactions with their environment. Hence, mass--size relations of galaxies have been studied to examine the formation mechanisms of various galaxies \citep{Shen2003,Bernardi2007,Hyde2009,Mosleh2013,Yoon2017,Kawinwanichakij2021}. For example, the mass--size relation for early type galaxies at $\log(M_\mathrm{star}/M_{\odot})>11.2$ shows an upwardly curved shape, deviating from the extrapolation of a linear relation for less-massive early type galaxies \citep{Desroches2007,Hyde2009,Bernardi2011b}. This curved relation at the more massive end can be explained by the increased role of major mergers in the formation of massive galaxies \citep{Bernardi2011a,Bernardi2011b}.

Furthermore, in order to understand how galaxy formation and evolution depend on environment, many previous studies have investigated mass--size relations of early type/quiescent galaxies in different environments (e.g., \citealt{Cooper2012,Lani2013,Delaye2014,Kelkar2015,Yoon2017}).\footnote{See Table 1 in \citet{Yoon2017} for more references.} For instance, \citet{Yoon2017} discovered the environmental dependence of the mass--size relation of early type galaxies at the most massive end with $\log(M_\mathrm{star}/M_{\odot})>11.2$, as expected in hierarchical galaxy simulations \citep{Shankar2013,Shankar2014}. \citet{Yoon2017} demonstrated that the most massive early type galaxies in dense environments have experienced more frequent galaxy mergers during their entire growth histories than those in less-dense environments. They suggest that this discrepancy in merger frequency is responsible for the environmental dependence of the mass--size relation of massive early type galaxies with $\log(M_\mathrm{star}/M_{\odot})>11.2$. This study shows that the different formations and evolutions that galaxies experience in various environments are able to cause differences in galaxy sizes for a given mass.

It is known that low-mass galaxies with $\log(M_\mathrm{star}/M_{\odot})\lesssim10.0$ in dense environments (e.g., clusters and groups) are strongly affected by environmental effects \citep{Haines2007,Peng2010,Wetzel2013,Moutard2018}, such as strangulation (i.e., starvation; \citealt{Larson1980}), ram pressure stripping \citep{Gunn1972}, and harassment \citep{Moore1996}, due to their shallow potential wells.\footnote{These environmental effects on low-mass galaxies are more efficient at lower redshifts \citep{Peng2010,Lee2015,Kawinwanichakij2017}.} These environmental effects are able to quench star formation activity efficiently, thereby producing quiescent galaxies from star-forming ones, and can also transform the morphologies of galaxies. Therefore, it is naturally expected that the slope of the mass--size relation of quiescent galaxies in the low-mass range of $\log(M_\mathrm{star}/M_{\odot})\lesssim10.0$ differs from that of the high-mass range, where the environmental effects are less dominant.

Indeed, several observation studies show that the slope of the mass--size relation of quiescent galaxies depends on mass in such a way that low-mass quiescent galaxies (hereafter, LQGs) with $\log(M_\mathrm{star}/M_{\odot})\lesssim10.0$ display a shallower slope  than high-mass quiescent galaxies (hereafter, HQGs) with $\log(M_\mathrm{star}/M_{\odot})\gtrsim10.5$ \citep{Mosleh2013,Lange2015,Kawinwanichakij2021,Nedkova2021}. So, the mass--size relation of quiescent galaxies is well fitted by a broken power-law function with a pivot mass of $\log(M_\mathrm{star}/M_{\odot})\sim10.3$ at low redshifts. 

\citet{Kawinwanichakij2021} show that the mass--size relation of LQGs has a shallow slope similar to that of low-mass star-forming galaxies. Based on this result, \citet{Kawinwanichakij2021} infer that LQGs with large sizes were recently drawn from star-forming galaxies through environmental quenching mechanisms, without significant morphological transformations. If this is true, we would expect LQGs with $\log(M_\mathrm{star}/M_{\odot})\lesssim10.0$ to have a shallower slope in the mass--size relation and therefore larger sizes in denser environments. However, no significant environmental dependence of the mass--size relation of LQGs has been found so far.

Using data from Sloan Digital Sky Survey (SDSS), we discover that the mass--size relation of quiescent galaxies with $\log(M_\mathrm{star}/M_{\odot})\lesssim10.0$ significantly depends on environment, suggesting that environmental effects play a crucial role in the formation of LQGs. This discovery is attributed to the large sample of galaxies that is complete down to a low stellar mass of $\log(M_\mathrm{star}/M_{\odot})\sim9.4$, which is essential for defining the environments of low-mass galaxies. We also examine the mass--size relation of LQGs in very underdense or isolated environments, where environmental effects in dense environments are negligible, and deduce their origins.

In this study, we use $H_0=70$ km s$^{-1}$ Mpc$^{-1}$, $\Omega_{\Lambda}=0.7$, and $\Omega_\mathrm{m}=0.3$ as cosmological parameters. 
\\

\begin{figure*}
\includegraphics[width=\linewidth]{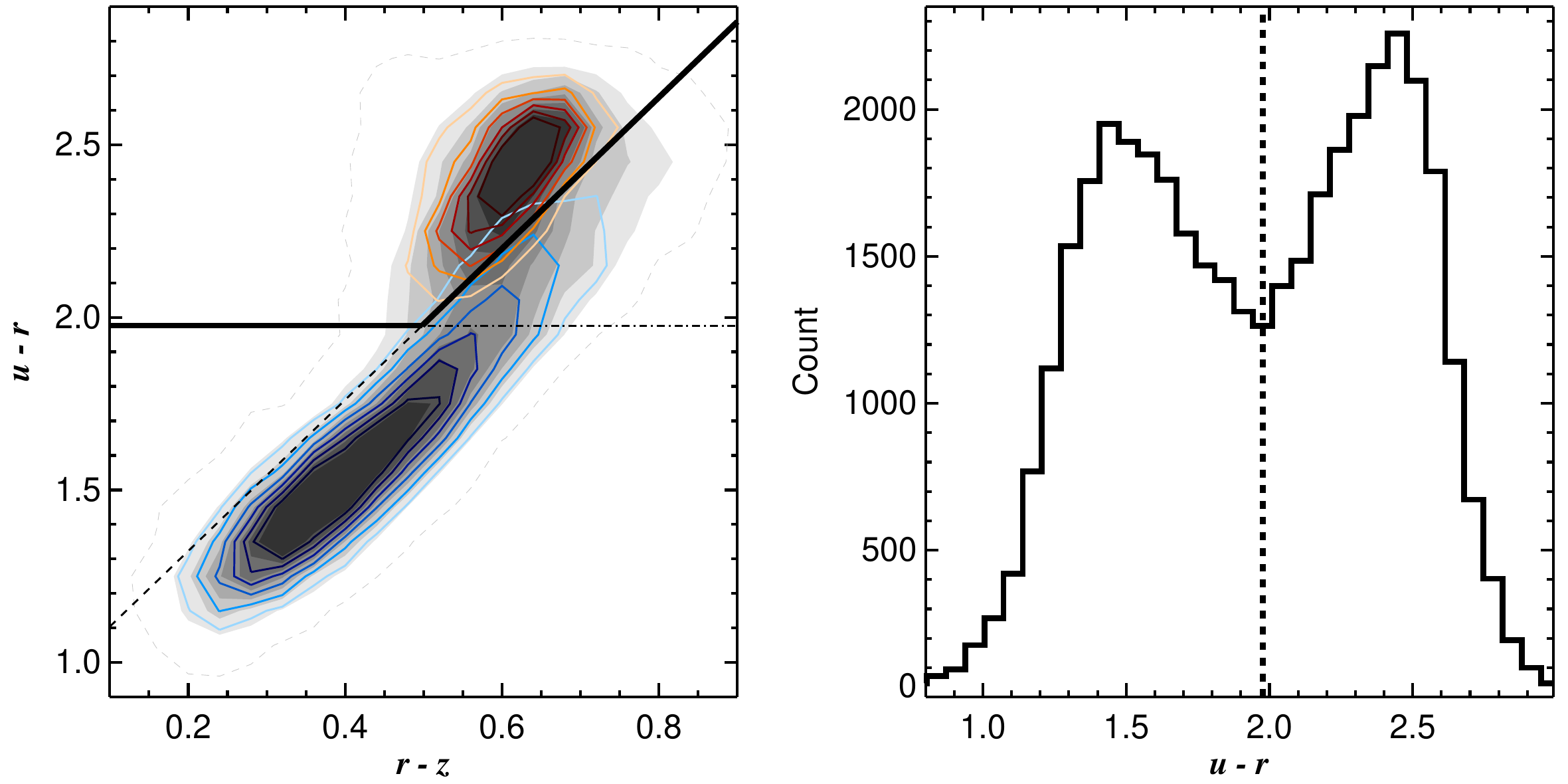}
\caption{Left panel: distribution of our sample in the $u-r$ versus $r-z$ color--color diagram. The filled gray contours indicate the densities of the data points (darker contours mean a higher density of data points). Note that $90\%$ of the galaxies in the sample is located within the outermost gray dashed contour line. The red contour lines denote the densities of the galaxies with EW(H$\alpha$)$\ge-1\mathrm{\AA}$, whereas the blue contour lines represent the densities of the EW(H$\alpha$)$<-1\mathrm{\AA}$ (an emission line has a negative value of EW(H$\alpha$)). The thick black solid line indicates the color cut that distinguishes between quiescent and star-forming galaxies (Equations (\ref{eq:cut1}) and (\ref{eq:cut2}); galaxies located in the upper region of the line are quiescent). The black dotted-dashed line represents Equation (\ref{eq:cut1}), while the black dashed line represents Equation (\ref{eq:cut2}). Right panel: histogram of the $u-r$ color distribution of our sample. The vertical dashed line indicates the local minimum between the two peaks in the $u-r$ color distribution, which corresponds to the horizontal cut in the $urz$ diagram (Equation (\ref{eq:cut1})).
\label{fig:cut}}
\end{figure*} 

\section{Sample}\label{sec:sample}

\subsection{Data and Catalogs}\label{sec:catalog}
The galaxy sample used in this study is extracted from the NASA-Sloan Atlas (NSA) catalog,\footnote{The details about the image analysis performed for the catalog are in \citet{Blanton2011}.} which includes virtually all known redshifts out to $z=0.15$ for galaxies within the coverage of SDSS data release (DR) 11 (the catalog contains about 640,000 galaxies).\footnote{The NSA catalog is also the base catalog for the Mapping Nearby Galaxies at Apache Point Observatory (MaNGA) survey project \citep{Wake2017}.} We focus on galaxies within the redshift range of $0.01<z<0.04$. The upper limit of the redshift range is set to create a volume-limited sample which is complete down to low stellar masses of $\log(M_\mathrm{star}/M_{\odot})\sim9.4$.

At the upper limit of $z=0.04$, the $r$-band magnitude limit for the spectroscopic target selection (the main galaxy sample; \citealt{Strauss2002}), which is $m_r\approx17.77$, corresponds to  $\log(M_\mathrm{star}/M_{\odot})\sim9.15$. At $z\sim0.04$, roughly $80\%$ of the galaxies with $m_r\approx17.77$ are at $\log(M_\mathrm{star}/M_{\odot})<9.4$, and more than $\sim90\%$ are at $\log(M_\mathrm{star}/M_{\odot})<9.5$. At $z\sim0.03$, which is the median redshift of our final sample, $m_r\approx17.77$ corresponds to $\log(M_\mathrm{star}/M_{\odot})\sim9.0$, and more than $\sim95\%$ of the galaxies with $m_r\approx17.77$ are at $\log(M_\mathrm{star}/M_{\odot})<9.4$. Thus, we set a stellar mass cut of $\log(M_\mathrm{star}/M_{\odot})\ge9.4$ in the sample. The number of galaxies in the volume-limited sample with $0.01<z<0.04$ and $\log(M_\mathrm{star}/M_{\odot})\ge9.4$ is 53,641 in the NSA catalog.

Stellar masses (and also magnitudes) used in this study are from two-dimensional S{\'e}rsic fits in the NSA catalog. The galaxy size used here is the half-light radius in the $r$ band ($R_e$), which is derived from the two-dimensional S{\'e}rsic fit. The half-light radius $R_e$ is obtained from the half-light semimajor axis length $a_e$ through $R_e=a_e\sqrt{q}$, in which $q$ is the axis ratio of the S{\'e}rsic model. While we use $R_e$ derived from the two-dimensional S{\'e}rsic fit to define the galaxy size throughout the paper, our results remain valid even if other definitions, such as the Petrosian $50\%$ and $90\%$ light radii, are used instead (see Appendix \ref{Appendix_A}). Similarly, our results remain unchanged by the use of alternative stellar masses from the Max Planck Institute for Astrophysics--Johns Hopkins University (MPA--JHU) catalog,\footnote{\url{http://www.sdss.org/dr17/spectro/galaxy_mpajhu/}} which are derived from the model magnitudes of SDSS.

To ensure robust measurements of galaxy environments, we use galaxies that are within regions (the area within a $0.5\degr$ radius centered on each galaxy) whose spectroscopic completeness is larger than $70\%$ for galaxies with $m_r<17.77$, and that are located at least $\sim0.3\degr$ away from the survey area boundary. We note that $0.5\degr$ corresponds to $\sim1.1$ Mpc at the median redshift of our final sample $z\sim0.03$. Such criteria based on spectroscopic completeness mostly select galaxies in the main galaxy sample survey areas \citep{Strauss2002} with a typical spectroscopic completeness of $\sim90\%$ for galaxies with $m_r\lesssim17.77$. The number of galaxies in the final sample is then 40,515. 

The average spectroscopic completeness around galaxies in our final sample is $90\%$, and $95\%$ of the galaxies in the sample are located in regions with spectroscopic completeness larger than $80\%$. We note that the use of stricter criteria (galaxies within regions with spectroscopic completeness larger than $80\%$ and located at least $\sim0.5\degr$ away from the boundary) does not alter our results, but only increases the statistical uncertainties due to the reduced number of galaxies in the final sample (37,672 galaxies).

The information of the H$\alpha$ equivalent width (EW(H$\alpha$)) and the median signal-to-noise ratio per pixel of the whole spectrum ((S/N)$_\mathrm{m}$) used in Section \ref{sec:quie} is from the MPA--JHU catalog, which is based on the spectroscopic data of SDSS DR8. We note that 36,810 galaxies in the MPA--JHU catalog match our final sample. 
\\

\subsection{Quiescent Galaxies}\label{sec:quie}
We classify quiescent galaxies using the $u-r$ versus $r-z$ color--color diagram ($urz$ diagram), which has been widely applied in previous studies to distinguish between quiescent and star-forming galaxies \citep{Holden2012,Robaina2012,McIntosh2014,Chang2015,Lacerna2016,Lopes2016,Kawinwanichakij2021}, with the aid of auxiliary information such as EW(H$\alpha$). The left panel of Figure \ref{fig:cut} displays the distribution of our sample in the $urz$ diagram, with the lines that distinguish quiescent galaxies from star-forming ones. The region of quiescent galaxies in the $urz$ diagram is defined as
\begin{eqnarray}
u-r &>&  1.976 \label{eq:cut1}\\
u-r &>&  2.189\times(r-z) + 0.885. \label{eq:cut2}
\end{eqnarray}

The horizontal cut in the $urz$ diagram (Equation (\ref{eq:cut1})) is set to correspond to the local minimum between the two peaks in the $u-r$ color distribution of our sample shown in the right panel of Figure \ref{fig:cut}. The diagonal boundary (Equation (\ref{eq:cut2})) is defined using EW(H$\alpha$) of galaxies with (S/N)$_\mathrm{m}>5$. We divide galaxies into two groups based on an EW(H$\alpha$) threshold of $-1\mathrm{\AA}$ (here, an emission line has a negative value of EW(H$\alpha$)) as in \citet{Ko2013}. Galaxies with EW(H$\alpha$)$\ge-1\mathrm{\AA}$ represent galaxies without star formation activity, whereas those with EW(H$\alpha$)$<-1\mathrm{\AA}$ correspond to galaxies with star formation activity. As shown in the left panel of Figure \ref{fig:cut}, the two groups occupy distinct regions on the $urz$ diagram: one group has a red $u-r$ color (the red contours), while the other group has a blue $u-r$ color (the blue contours). The latter group forms a relatively long sequence in the diagram. The slope of the diagonal line (2.189) is derived by fitting a line (minimum $\chi^2$ fit) to the galaxies with EW(H$\alpha$)$<-1\mathrm{\AA}$ with clipping $3\sigma$ outliers in the fitting process. After fixing the slope, we determine the zero-point of the line that satisfies the condition $f_c=f_r$. In this condition, $f_c$ means the fraction of galaxies classified as quiescent galaxies by the color cut (Equations (\ref{eq:cut1}) and (\ref{eq:cut2})), among the galaxies with EW(H$\alpha$)$\ge-1\mathrm{\AA}$. The parameter $f_r$ indicates the fraction of galaxies with EW(H$\alpha$)$\ge-1\mathrm{\AA}$, among the galaxies classified as quiescent galaxies by the color cut. Thus, $f_c$ is a concept similar to completeness, whereas $f_r$ is a concept related to reliability. We find that both $f_c$ and $f_r$ are 0.824 simultaneously when the zero-point of the diagonal line is 0.885. The number of quiescent galaxies defined by the color cut is 13,667.

We use the color cut of Equations (\ref{eq:cut1}) and (\ref{eq:cut2}) to define quiescent galaxies throughout the paper. However, our results are essentially identical if we use an EW(H$\alpha$) threshold of $-1\mathrm{\AA}$ as the main criterion to separate quiescent galaxies from star-forming galaxies, as shown in Appendix \ref{Appendix_A}. There are two reasons why we use the color cut instead of EW(H$\alpha$). First, the EWs used here only represent the central regions ($<1.5\arcsec$) of galaxies, since the diameter of the SDSS fiber is $3\arcsec$. Second, galaxy color is a more easily accessible quantity, and as such, it has been widely used in previous studies to classify quiescent galaxies and will also be frequently used in future studies.
\\

\begin{figure}
\includegraphics[width=\linewidth]{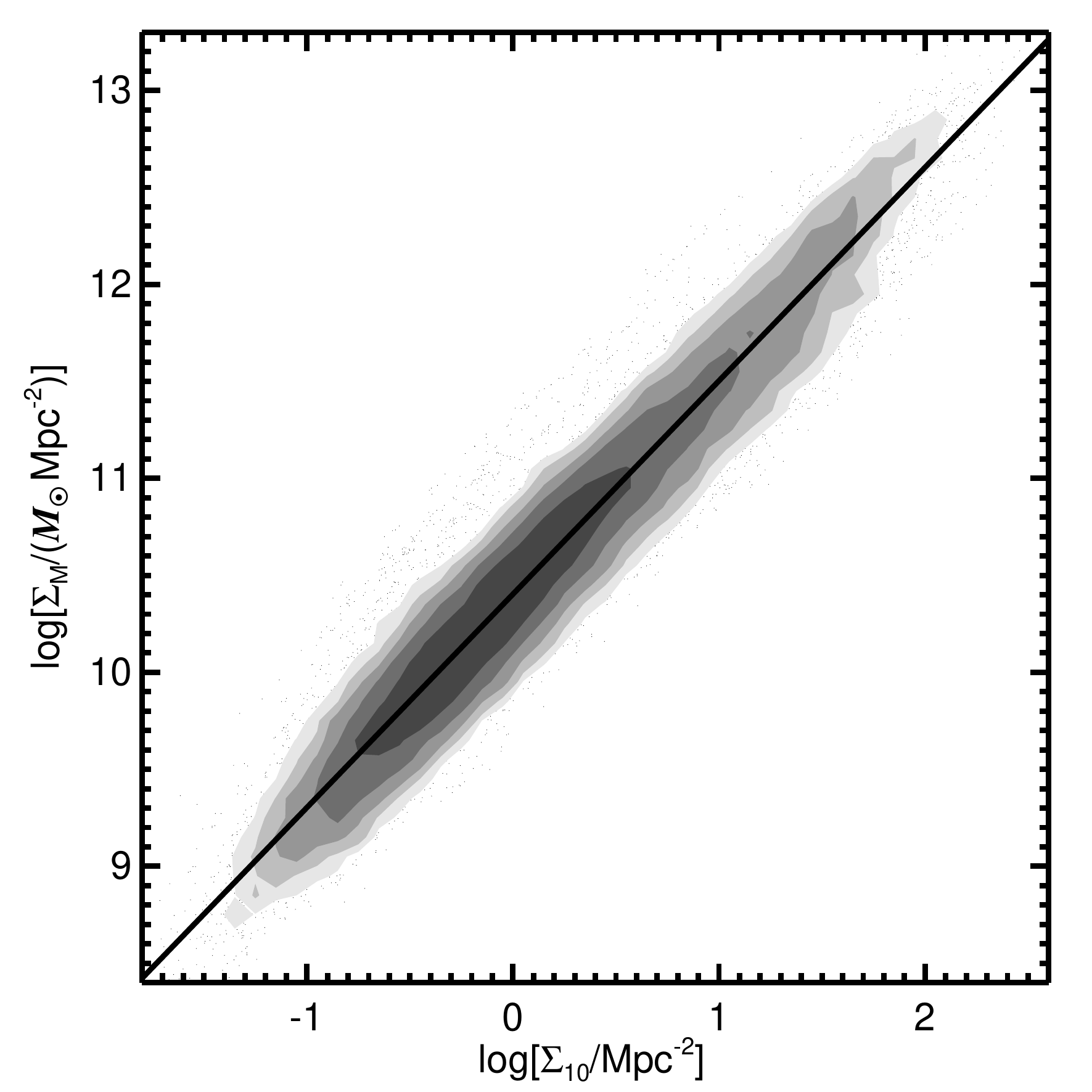}
\caption{Comparison between the logarithmic values of  $\Sigma_{10}$ and $\Sigma_{M}$. The filled gray contours indicate the density of data points (darker contours mean a higher density of data points). The black solid line indicates the line fitted to the distribution of data points. The equation of the line is $\log\Sigma_{M} = 1.1\log\Sigma_{10}+10.4$. In this figure, we only show data points within $5\sigma$ from the line, where $\sigma$ is 0.2 dex in the direction of the $y$-axis. The two quantities show a nearly perfect correlation with a Spearman's correlation coefficient of 0.95 and a small scatter of 0.2 dex. Therefore, they are almost interchangeable.
\label{fig:dencomp}}
\end{figure} 

\begin{figure}
\includegraphics[width=\linewidth]{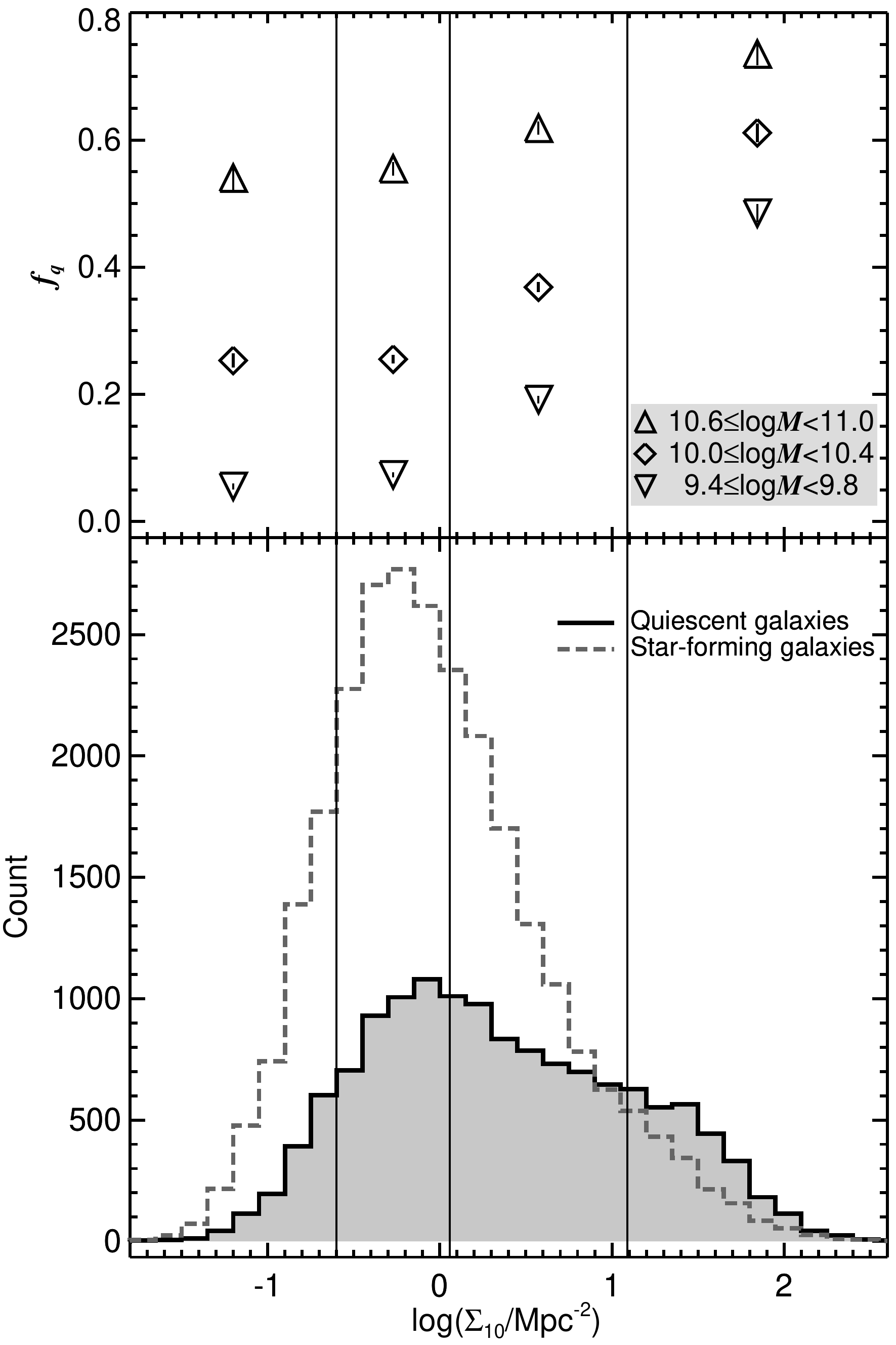}
\caption{Lower panel: surface number density ($\log\Sigma_{10}$) distributions for quiescent (the black solid line)  and star-forming galaxies (the gray dashed line). The vertical lines in both panels divide the quiescent galaxies into the four environment bins used in this study. Upper panel: quiescent fractions $f_q$ for galaxies with different stellar masses in the four environment bins.
\label{fig:dendist}}
\end{figure}

\begin{figure}
\includegraphics[width=\linewidth]{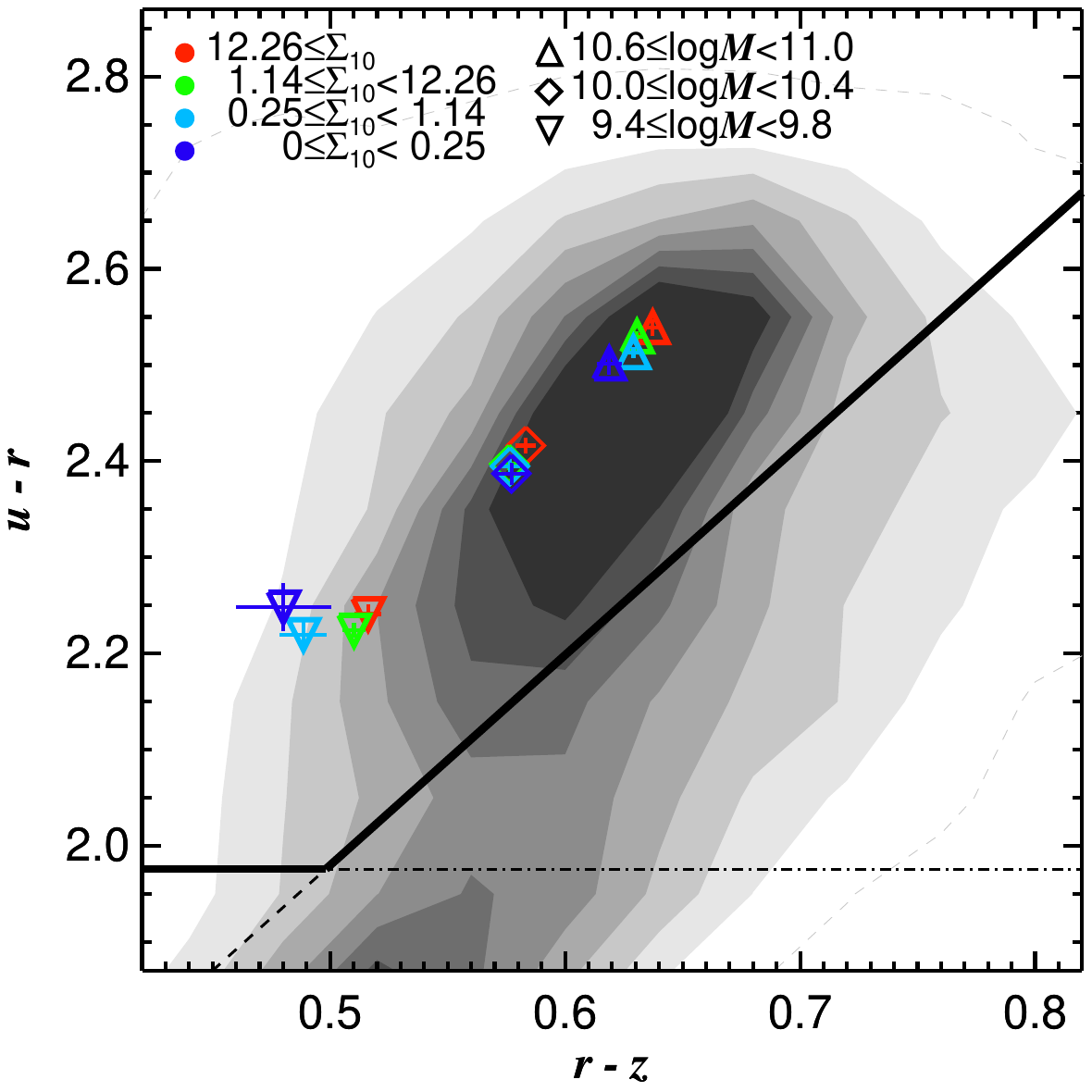}
\caption{Quiescent region of the $urz$ diagram from Figure \ref{fig:cut}, in which the median color of quiescent galaxies with different stellar masses in the four environment bins is plotted on the diagram. The data points are differentiated by shape to indicate the different stellar mass bins and by colors to represent the different environment bins. The error bars denote the standard error of the median. The units of $\Sigma_{10}$ in the legend are Mpc$^{-2}$.
\label{fig:cs}}
\end{figure}

\begin{figure*}
\includegraphics[width=\linewidth]{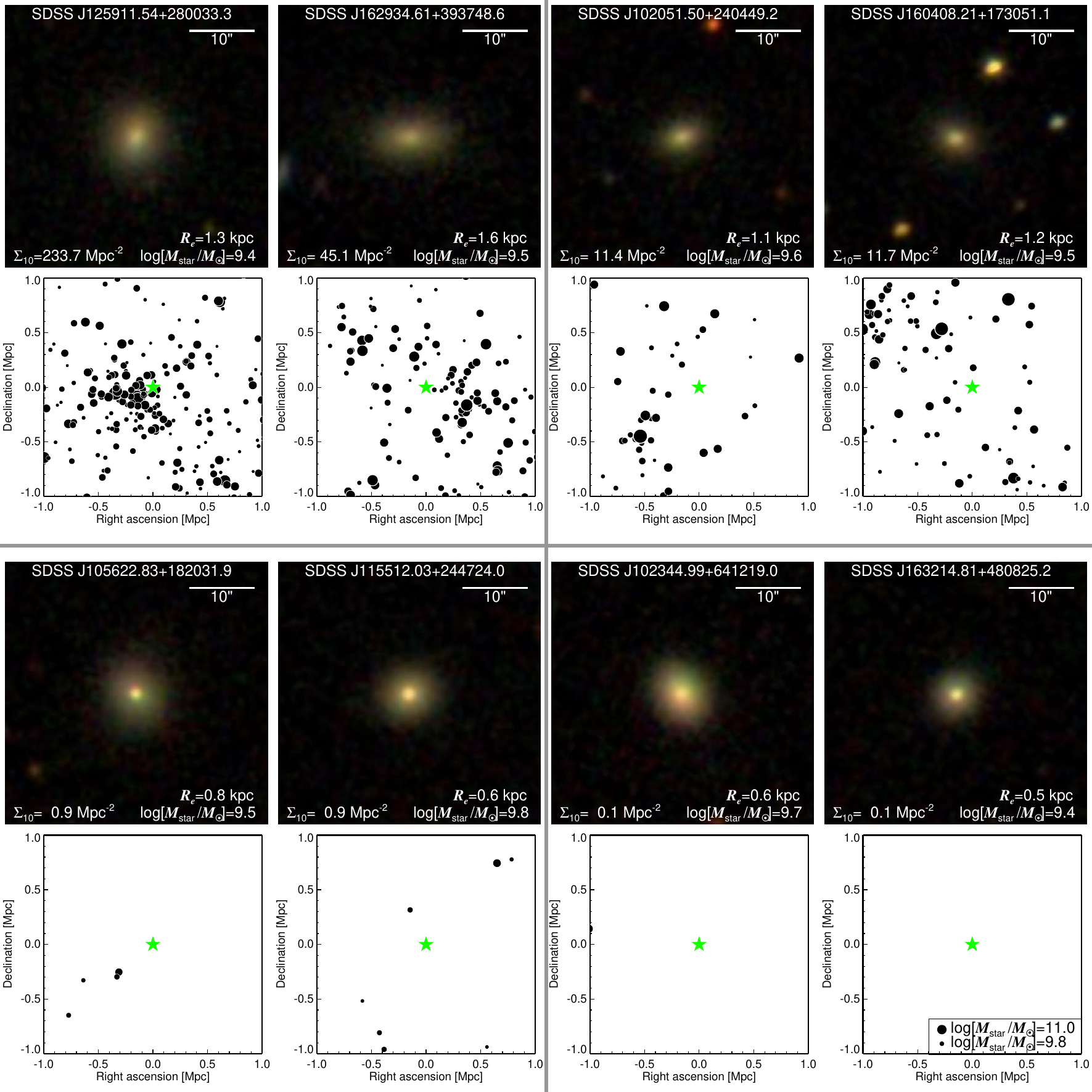}
\caption{Color images of LQGs with $\log(M_\mathrm{star}/M_{\odot})\leq9.8$ and spatial distribution map for galaxies within $\sim1$ Mpc of each LQG. By the gray separation lines, this figure is divided into four parts, which represent galaxies in the four environments bins used in this study. For example, the upper left part is for galaxies in the highest-density bin, while the lower right part is for galaxies in the lowest-density bin. The galaxy ID, $R_e$, $\Sigma_{10}$, and stellar mass are shown in the color image. The horizontal bar in the color image indicates the angular scale of the image. The green star in the galaxy map indicates the location of the LQG, while the black filled circles represent other galaxies in the environment. As indicated in the accompanying legend in the bottom right panel, the size of the circle denotes the stellar mass of the galaxy.
\label{fig:ex1}}
\end{figure*}

\begin{figure*}
\includegraphics[width=\linewidth]{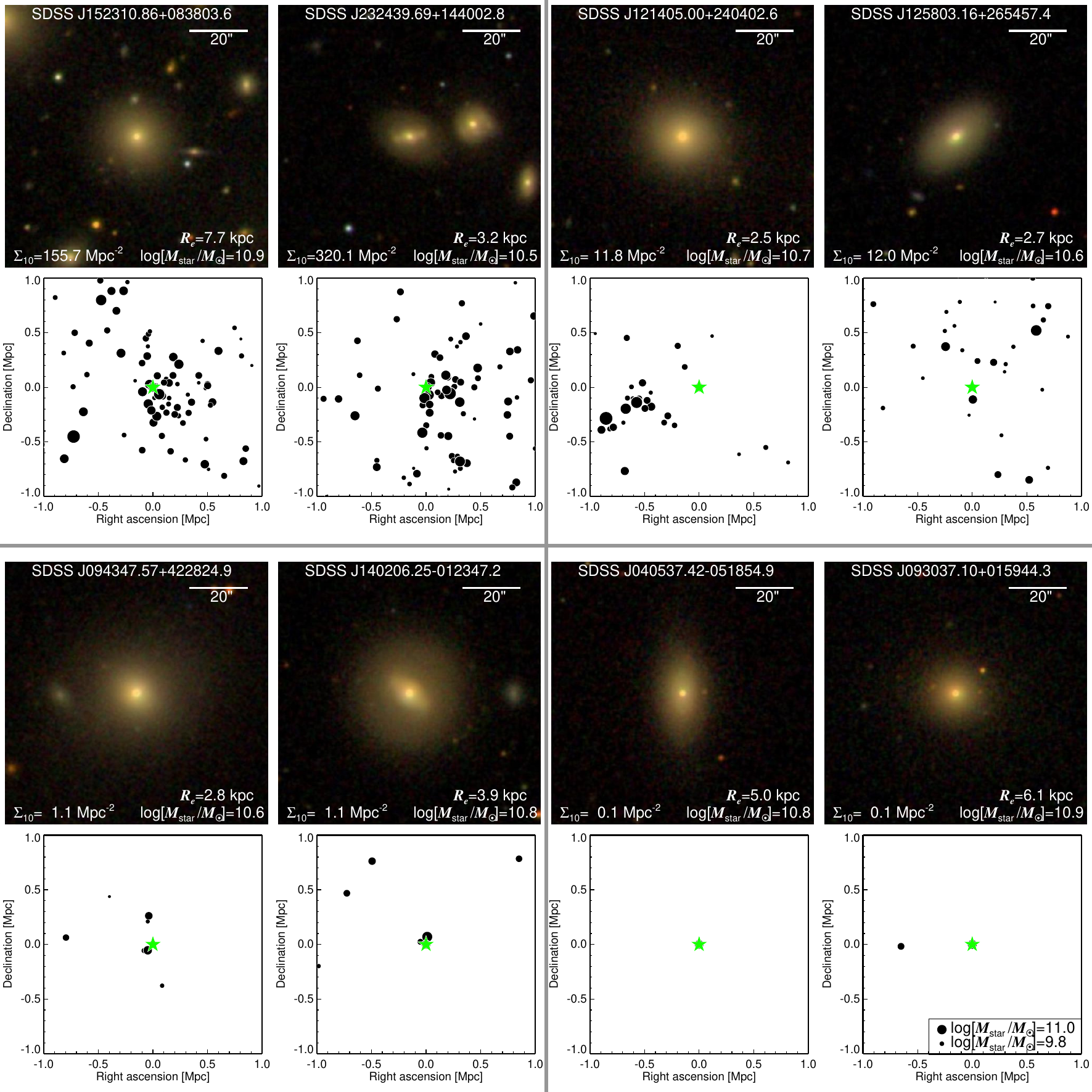}
\caption{Color images of HQGs with $\log(M_\mathrm{star}/M_{\odot})\geq10.5$ and spatial distribution map for galaxies within $\sim1$ Mpc of each HQG. The other details for this figure are the same as those in Figure \ref{fig:ex1}.
\label{fig:ex2}}
\end{figure*}

\subsection{Environments of Galaxies}\label{sec:env}
In this study, we define the environment of a galaxy as the surface number density of galaxies within the distance to the tenth-nearest galaxy. The surface number density is computed using galaxies with $\log(M_\mathrm{star}/M_{\odot})\ge9.4$ within a rest-frame velocity range of $\pm1000$ km s$^{-1}$ centered on the redshift of a galaxy in our sample.\footnote{Thus, the redshift range of galaxies for calculating environments should be slightly broader ($0.006<z<0.044$) than that of the galaxy sample.} The surface number density is then calculated by
\begin{equation}
\Sigma_{10}=\frac{10}{\pi {r_{10}}^2},
\label{eq:den}
\end{equation}
where $r_{10}$ is projected distance to the tenth-nearest neighbor (the units of $r_{10}$ are Mpc in this study). As an alternative environment, the surface stellar mass density using the ten nearest neighbor galaxies is computed by 
\begin{equation}
\Sigma_{M}=\frac{\Sigma_{i=1}^{10} M_{\mathrm{star}, i}}{\pi {r_{10}}^2},
\label{eq:denm}
\end{equation}
where $M_{\mathrm{star}, i}$ is the stellar mass of the $i$th galaxy.

We compare $\Sigma_{10}$ with $\Sigma_{M}$ in Figure \ref{fig:dencomp}, which shows that the two quantities display a nearly perfect correlation.\footnote{This implies that the shape of the stellar mass function of galaxies is consistent across various environments.} The Spearman's correlation coefficient between $\Sigma_{10}$ and $\Sigma_{M}$ is 0.95, and the scatter of the data points around the relation between $\Sigma_{10}$ and $\Sigma_{M}$ (see the caption of Figure \ref{fig:dencomp}) is only 0.2 dex. We find that $\Sigma_{10}$ and $\Sigma_{M}$ are interchangeable, resulting in nearly identical outcomes based on either quantity, as shown in Appendix \ref{Appendix_A}. In this study, we choose to present our results based on $\Sigma_{10}$ due to its advantage of simplicity, which makes it easily applicable to other data sets.

The lower panel of Figure \ref{fig:dendist} displays the surface number density distributions for quiescent and star-forming galaxies. We divide quiescent galaxies into four environment bins: $0\leq\Sigma_{10}/\mathrm{Mpc}^{-2}<0.25$, $0.25\leq\Sigma_{10}/\mathrm{Mpc}^{-2}<1.14$, $1.14\leq\Sigma_{10}/\mathrm{Mpc}^{-2}<12.26$, and $\Sigma_{10}/\mathrm{Mpc}^{-2}\ge12.26$. These four bins correspond to percentile ranges of 0 -- 10\%, 10 -- 40\%, 40 -- 80\%, and 80 -- 100\%, respectively, for the environments of quiescent galaxies.

In order to investigate the basic properties of the four environment bins, we examine the quiescent fraction $f_q$\footnote{The quiescent fraction is calculated as the number of quiescent galaxies divided by the total number of galaxies.} as a function of $\log\Sigma_{10}$, which is displayed in the upper panel of Figure \ref{fig:dendist}. The fraction $f_q$ of high-mass galaxies with $10.6\leq\log(M_\mathrm{star}/M_{\odot})<11.0$ is higher than 0.5 in all the $\Sigma_{10}$ bins. In the highest-density bin, $74\%$ of the high-mass galaxies are quiescent. By contrast, only a very small fraction ($5.5\%$) of low-mass galaxies with $\log(M_\mathrm{star}/M_{\odot})<9.8$ are quiescent in the lowest-density environments, while nearly half of the low-mass galaxies ($48\%$) are quiescent in the highest-density environments. Thus, the $f_q$ of low-mass galaxies varies more significantly with environment than that of high-mass galaxies.

We also examine the typical color of quiescent galaxies in each environment. Displayed in Figure \ref{fig:cs} is the quiescent region of the $urz$ diagram from Figure \ref{fig:cut}, where the median color of quiescent galaxies with different stellar masses in the four environment bins is plotted. According to the figure, the color of quiescent galaxies appears to be far less dependent on the environment, with stellar mass being the primary factor driving color variations (galaxies with higher stellar masses are redder in both colors). Although there is a minor trend that quiescent galaxies in higher-density environments are slightly redder, the differences in color between the environment bins are minimal, with a maximum variation of only 0.04. We note that the slight variations in galaxy color across environments do not result in significant differences in galaxy sizes between the different environment bins, as described in Appendix \ref{Appendix_B}. Thus, color is not a critical factor driving our main findings shown in Section \ref{sec:results}.

Figures \ref{fig:ex1} and \ref{fig:ex2} display color images of quiescent galaxies and spatial distribution maps for galaxies within $\sim1$ Mpc of each quiescent galaxy (Figure \ref{fig:ex1} is for LQGs with $\log(M_\mathrm{star}/M_{\odot})\leq9.8$, whereas Figure \ref{fig:ex2} is for high-mass counterparts with $\log(M_\mathrm{star}/M_{\odot})\geq10.5$). The spatial distribution maps shown in the two figures illustrate that the highest-density bin of $\Sigma_{10}/\mathrm{Mpc}^{-2}\ge12.26$ represents very dense environments, such as clusters, with typically tens to hundreds of galaxies within a radius of $\sim1$ Mpc. In contrast, the environment bin of $1.14\leq\Sigma_{10}/\mathrm{Mpc}^{-2}<12.26$ corresponds to regions outside of dense environments or group-scale environments, typically containing a few dozen galaxies within a radius of $\sim1$ Mpc.  Similarly, the environment bin of $0.25\leq\Sigma_{10}/\mathrm{Mpc}^{-2}<1.14$ represents low-density groups or field environments, characterized by a low galaxy density of a few galaxies within a radius of $\sim1$ Mpc. Finally, the lowest-density bin of $\Sigma_{10}/\mathrm{Mpc}^{-2}<0.25$ denotes (nearly) isolated environments with an extremely low galaxy density, where there are almost no galaxies located within a radius of $\sim1$ Mpc. Thus, this lowest-density bin is also referred to as an isolated environment in this study, although it should be noted that some environments within this bin are not perfectly isolated.
\\

\begin{figure}
\includegraphics[width=\linewidth]{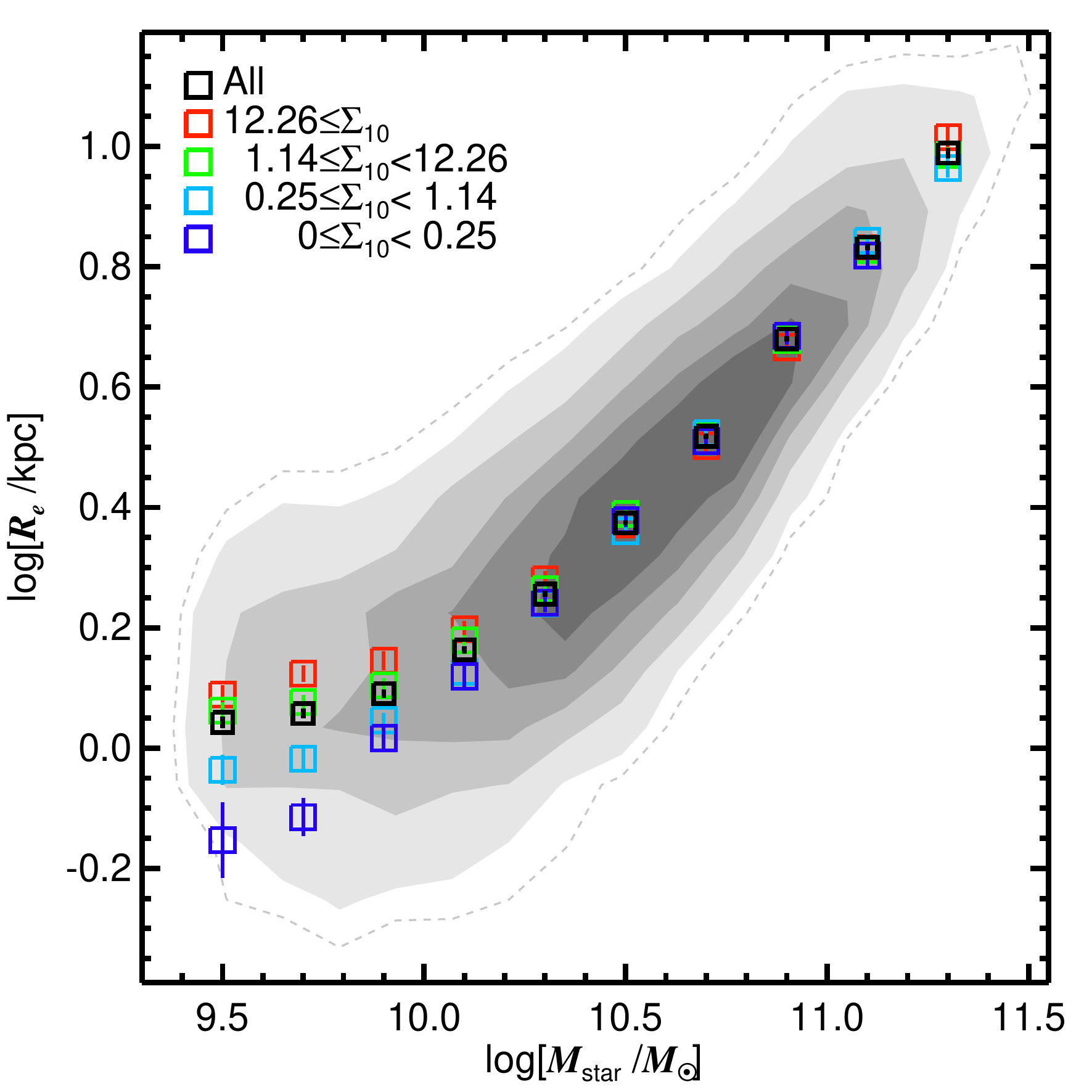}
\caption{Mass--size relation of our quiescent galaxies. The squares of the different colors represent the different environment bins, while the black squares denote all quiescent galaxies. A square indicates the mean $\log R_e$ of the mass bin (with a bin size of 0.2 dex), after excluding outliers beyond $5\sigma$. The error bars denote the standard error of the mean. This figure displays only bins with a minimum of 45 galaxies. The filled gray contours indicate the density of data points (a darker contour means a higher density of data points). Note that $90\%$ of the galaxies in the sample are located within the outermost gray dashed contour line. The units of $\Sigma_{10}$ in the legend are Mpc$^{-2}$.
\label{fig:ms}}
\end{figure} 

\begin{figure}
\includegraphics[width=\linewidth]{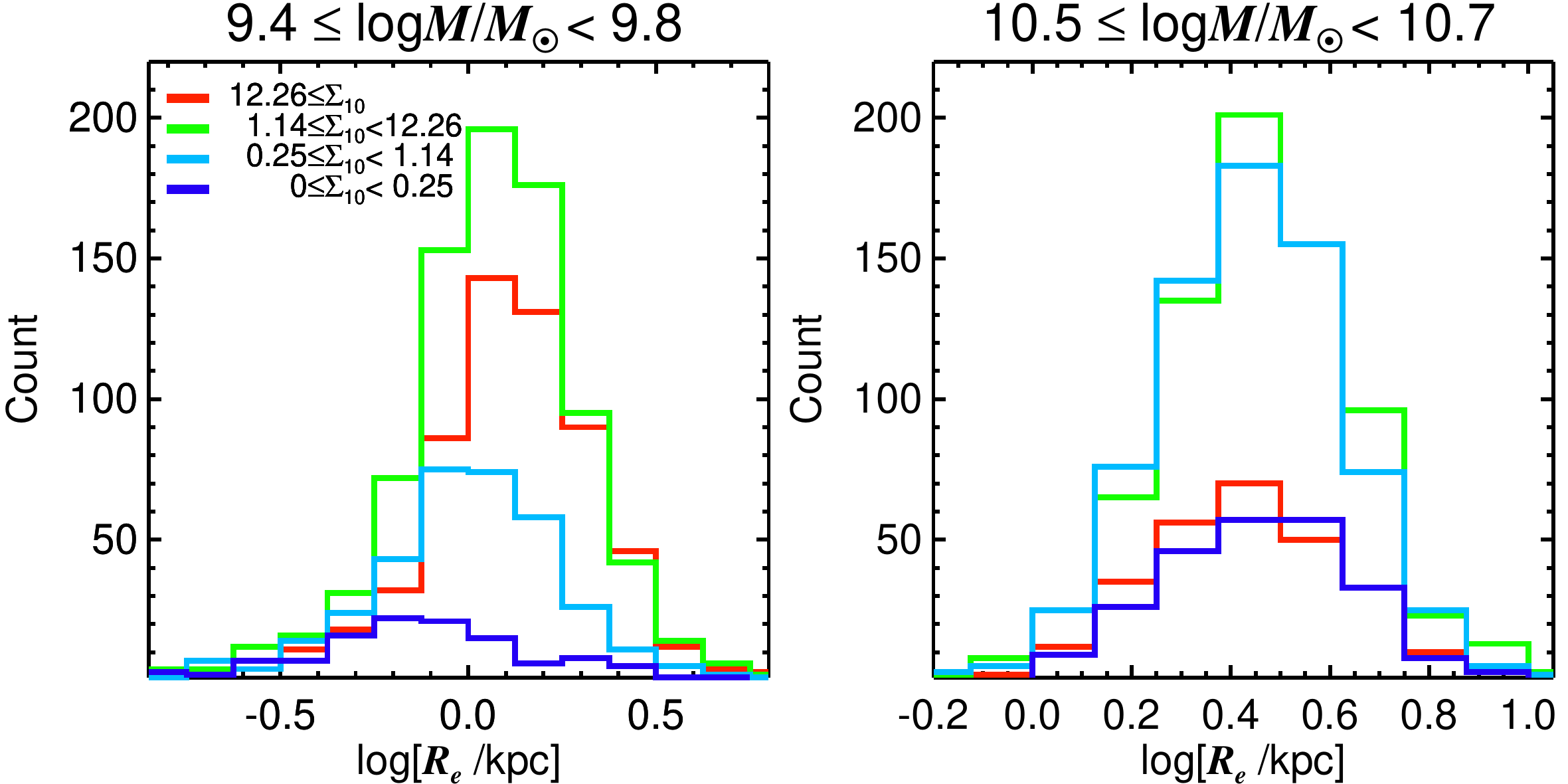}
\caption{Distribution of $\log R_e$ for LQGs with $9.4\le\log(M_\mathrm{star}/M_{\odot})<9.8$ (the left panel) and HQGs with $10.5\le\log(M_\mathrm{star}/M_{\odot})<10.7$ (the right panel). The different colors represent the different environment bins. The units of $\Sigma_{10}$ in the legend are Mpc$^{-2}$.
\label{fig:sizedist}}
\end{figure} 

\begin{figure}
\includegraphics[width=\linewidth]{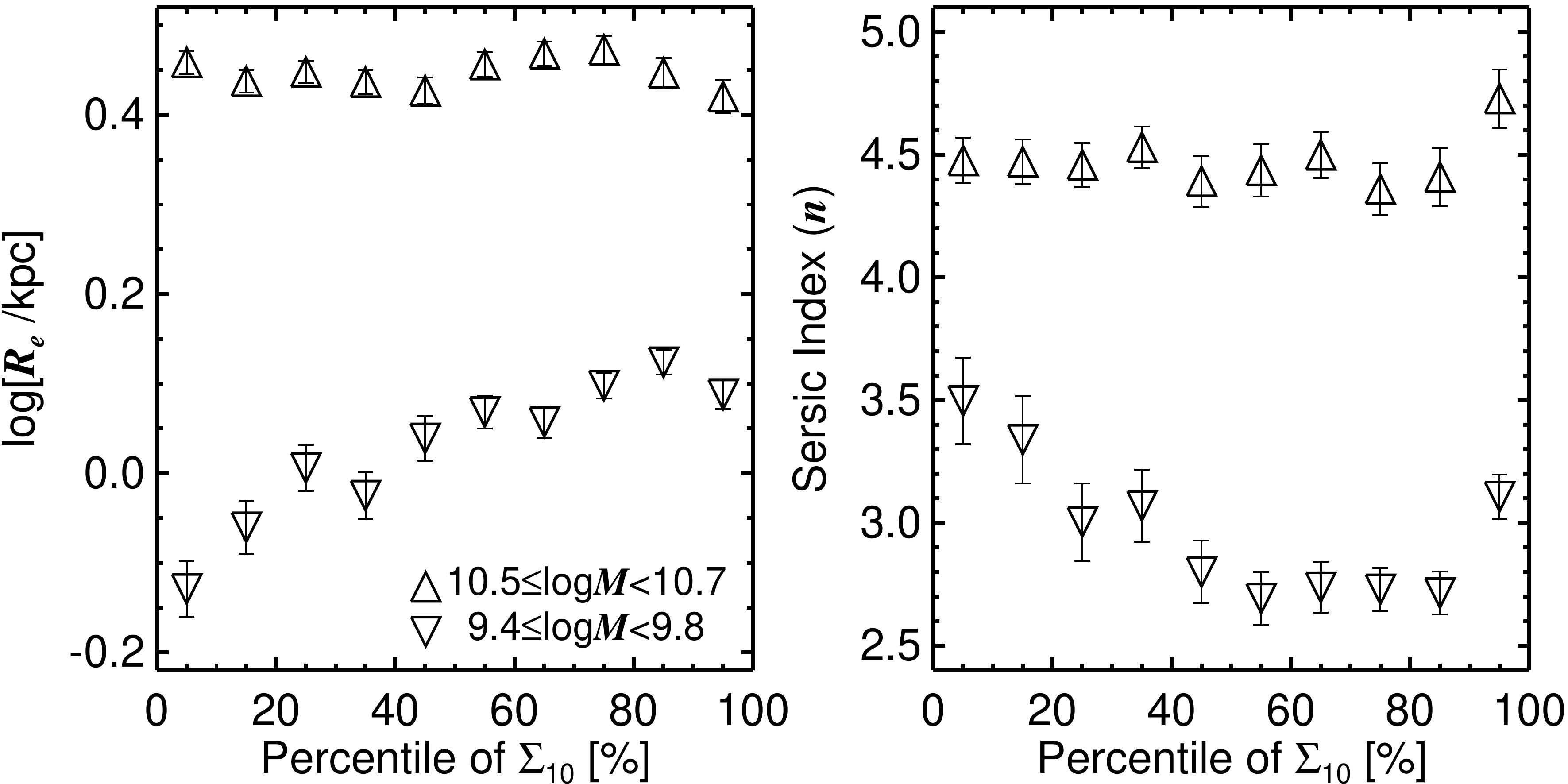}
\caption{Mean $R_e$ and S{\'e}rsic index ($n$) as a function of environment, with the exclusion of outliers beyond $5\sigma$ in the calculation of the mean value. Low percentiles correspond to low-density environments. The data points are differentiated by shape to indicate the different stellar mass bins. The error bars indicate the standard error of the mean. 
\label{fig:densize}}
\end{figure}

\begin{figure*}
\includegraphics[width=\linewidth]{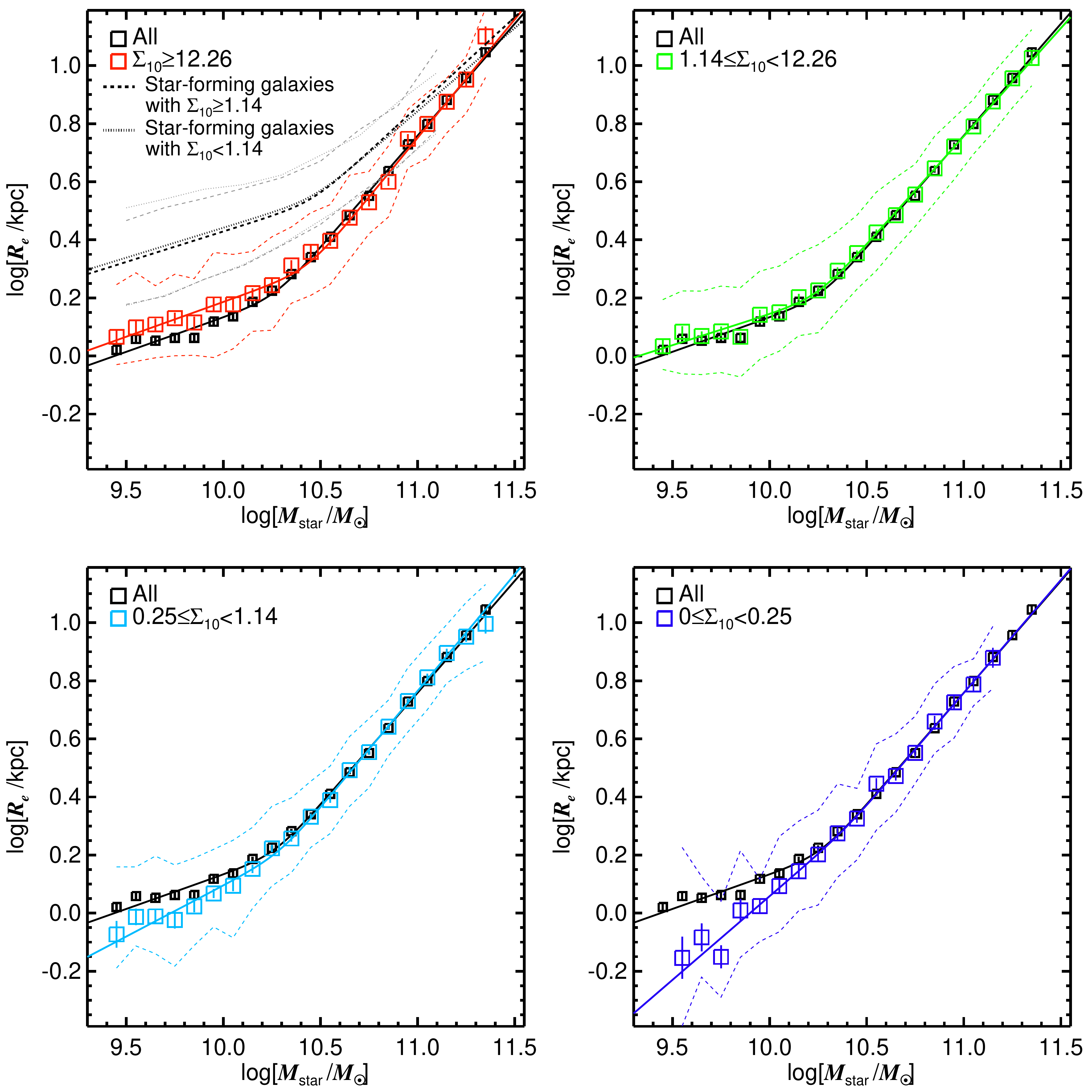}
\caption{Mass--size relations of quiescent galaxies in different environments with the fitted broken power-law functions of Equation (\ref{eq:bp}) overlaid on the relations. The mass--size relations of the different environment bins are displayed in the separate panels, each with its own color. A square indicates the mean $\log R_e$ of the mass bin (with a bin size of 0.1 dex), after excluding outliers beyond $5\sigma$. The error bars indicate the standard error of the mean. This figure shows only bins with a minimum of 25 galaxies. The colored dashed lines in each panel represent the 25th and 75th percentiles of $R_e$. The colored solid line in each panel indicates the smoothly broken power-law function fitted to each mass--size relation. The mass--size relation of all quiescent galaxies and its fitting result are shown in all the panels as the black squares and the black solid lines. In the upper left panel, we also display the smoothly broken power-law function (the black dashed line or dotted line) fitted to the mass--size relation of star-forming galaxies with $\Sigma_{10}/\mathrm{Mpc}^{-2}\ge1.14$ and $\Sigma_{10}/\mathrm{Mpc}^{-2}<1.14$, for the purpose of comparison. The gray dashed (or dotted) lines in the upper left panel represent the 25th and 75th percentiles of $R_e$ for star-forming galaxies. The units of $\Sigma_{10}$ in the legends are Mpc$^{-2}$.
\label{fig:msc}}
\end{figure*}

\section{Results}\label{sec:results}

Figure \ref{fig:ms} displays the mass--size relations of our quiescent galaxies in the different environment bins. As found in previous studies \citep{Mosleh2013,Lange2015,Kawinwanichakij2021,Nedkova2021}, the mass--size relation of LQGs with $\log(M_\mathrm{star}/M_{\odot})\lesssim10.0$ exhibits a shallower slope than that of HQGs with  $\log(M_\mathrm{star}/M_{\odot})\gtrsim10.5$, with a slope change occurring at $\log(M_\mathrm{star}/M_{\odot})\sim10.3$.  We find an interesting trend that the slope of the mass--size relation for LQGs is shallower in higher-density environments, while the slope of HQGs remains nearly identical across all environments. In other words, LQGs with $\log(M_\mathrm{star}/M_{\odot})\lesssim10.0$ are smaller in lower-density environments, while the sizes of HQGs with $\log(M_\mathrm{star}/M_{\odot})\gtrsim10.5$ do not depend on environment for a given mass. This trend can also be confirmed in Figure \ref{fig:sizedist}, which shows the distribution of $\log R_e$ for LQGs and HQGs in different environments, and in the left panel of Figure \ref{fig:densize}, which exhibits the mean $\log R_e$ as a function of environment for LQGs and HQGs. 

According to the figures, the mean $R_e$ for LQGs with $9.4\le\log(M_\mathrm{star}/M_{\odot})<9.8$ differs by 0.23 dex between the highest- and lowest-density environments, indicating that the sizes of LQGs in the highest-density bin are approximately $70\%$ larger on average than those in the lowest-density bin. The decrease in the sizes of LQGs in lower-density environments is a gradual phenomenon, as shown in the left panel of Figure \ref{fig:densize}.\footnote{It is known that bulge sizes smaller than $80\%$ of the half width at half maximum (HWHM) of the seeing can be overestimated \citep{Gadotti2008,Meert2015}. The typical HWHM of the seeing for the $r$ band is $0.66\arcsec$. Therefore, $80\%$ of the HWHM corresponds to $\log(R_e/\mathrm{kpc})=-0.38$ at $z=0.04$. As a result, most of our quiescent galaxies are not largely affected by the seeing effect. Moreover, the bias from the seeing effect can only weaken our main result that LQGs are smaller in lower-density environments, rather than artificially generating or enhancing the result.} By contrast, the mean $R_e$ values for HQGs with $10.5\le\log(M_\mathrm{star}/M_{\odot})<10.7$ in the different environment bins are consistent within 0.05 dex. 

To assess the significance of the size difference between LQGs with $9.4\le\log(M_\mathrm{star}/M_{\odot})<9.8$ in the four environment bins, Kolmogorov-Smirnov (K-S) tests are conducted on the size distributions in the left panel of Figure \ref{fig:sizedist}. The probability $(0\le p \le1)$ of the null hypothesis, in which the two sizes distributions for the highest- and lowest-density bins are drawn from the same distribution, is found to be $p=6.8\times10^{-17}$, indicating that the difference between the two size distributions is highly significant. We note that all $p$-values for the size distributions between any two environment bins of LQGs are less than $9\times10^{-4}$.

As shown in Appendix \ref{Appendix_A}, our main results on the environmental dependence of the mass--size relation do not essentially change if we use alternative definitions of quiescent galaxies (based on EW(H$\alpha$) or combining color with S{\'e}rsic indices), sizes (Petrosian $90\%$ light radii), or environments (surface stellar mass density). 

The right panel of Figure \ref{fig:densize} shows how the S{\'e}rsic index ($n$) of quiescent galaxies depends on environment for LQGs with $9.4\le\log(M_\mathrm{star}/M_{\odot})<9.8$ and HQGs with $10.5\le\log(M_\mathrm{star}/M_{\odot})<10.7$. The figure illustrates that the S{\'e}rsic index of LQGs clearly depends on environment, such that the mean S{\'e}rsic index of LQGs in the low-density environments is $n\approx3.5$, while that in the mid- to high-density environments is $n\approx2.7$. This implies that the structures of LQGs are relatively closer to classical bulges or elliptical galaxies ($n\sim4$) in lower-density environments. By contrast, the mean S{\'e}rsic index of HQGs is almost constant at $n\approx4.5$ across all the environments.

To scrutinize the slope of the mass--size relation for quiescent galaxies quantitatively, we fit smoothly broken power-law functions to the mass--size relations in different environments, as in \citet{Mowla2019} and \citet{Kawinwanichakij2021}. The smoothly broken power-law function is defined as
\begin{equation}
R_e(M_\mathrm{star})=r_p\biggl(\frac{M_\mathrm{star}}{M_p}\biggr)^\alpha \Biggl[\frac{1}{2}\Biggl\{1+ \biggl(\frac{M_\mathrm{star}}{M_p}\biggr)^\delta \Biggr\}\Biggr]^{(\beta-\alpha)/\delta}.
\label{eq:bp}
\end{equation}
In this function, $M_p$ is the pivot stellar mass where the slope of the function changes and $r_p$ is the radius at $M_p$. The parameter $\alpha$ indicates the power-law slope at the low-mass range at $M_\mathrm{star}\ll M_p$, whereas $\beta$ represents the slope at the high-mass range at $M_\mathrm{star}\gg M_p$. The parameter $\delta$ acts as a smoothing factor that determines the rate at which the slope of the function changes around the pivot stellar mass $M_p$, controlling the sharpness of the transition between the low-mass and high-mass ranges. We set $\delta=6$ following \citet{Mowla2019} and \citet{Kawinwanichakij2021}, in order to reduce the degeneracy between $\delta$ and the power-law slopes. We fit Equation (\ref{eq:bp}) to the mass--size relations of quiescent galaxies in the different environments using the minimum $\chi^2$ method. The fitting is performed in the stellar mass range of $\log(M_\mathrm{star}/M_{\odot})<11.3$ because another environmental dependence of the mass--size relation might emerge at the very massive end, as shown in \citet{Yoon2017}.

\begin{figure*}
\includegraphics[width=\linewidth]{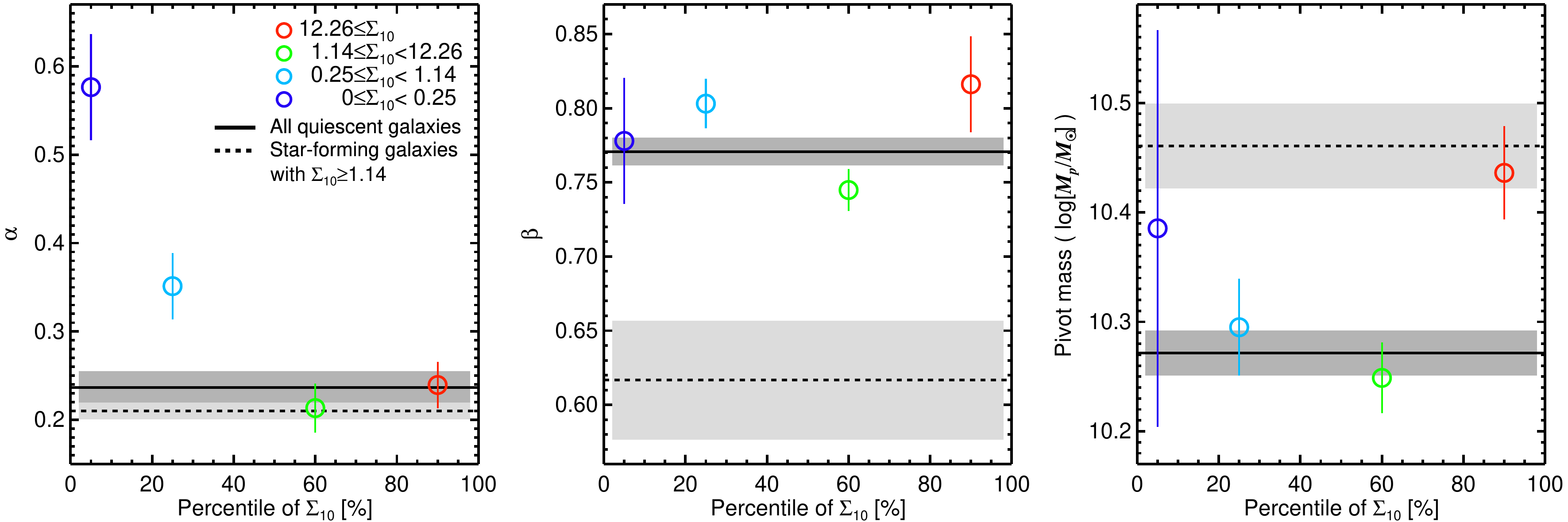}
\caption{Best-fit parameters ($\alpha$, $\beta$, and $\log M_p$) for the smoothly broken power-law functions (Equation (\ref{eq:bp})) for the different environment bins. The circles with different colors represent the parameters of the different environment bins. The black horizontal line with the dark gray area in each panel represent the best-fit parameter with its error for all quiescent galaxies. Similarly, the black dashed line with the light gray area indicates the best-fit parameter with its error for star-forming galaxies with $\Sigma_{10}/\mathrm{Mpc}^{-2}\ge1.14$. The units of $\Sigma_{10}$ in the legend are Mpc$^{-2}$.
\label{fig:coeff}}
\end{figure*}

 Figure \ref{fig:msc} displays the mass--size relations of quiescent galaxies in the different environment bins as well as the fitted broken power-law functions overlaid on the relations. Figure \ref{fig:coeff} illustrates the best-fit parameters for the broken power-law functions in the different environments. The values of these parameters are also tabulated in Table \ref{tb:best}. To evaluate any potential differences or similarities between the mass--size relations of quiescent and star-forming galaxies, we also show the fitting result for the mass--size relation of star-forming galaxies with $\Sigma_{10}/\mathrm{Mpc}^{-2}\ge1.14$ and $\Sigma_{10}/\mathrm{Mpc}^{-2}<1.14$ in Figures \ref{fig:msc} and \ref{fig:coeff}, and Table \ref{tb:best}.\footnote{We find that the mass--size relation of star-forming galaxies depends very little on environment (see Figure \ref{fig:msc} and Table \ref{tb:best}).}

Figures \ref{fig:msc} and \ref{fig:coeff} show that the slopes at the high-mass range are nearly constant within their errors across all environment bins ($\beta\approx0.77$). On the other hand, the slope at the low-mass range ($\alpha$) is steeper in the lower-density environments, so that the $\alpha$ of the high-density environment of $\Sigma_{10}/\mathrm{Mpc}^{-2}\ge1.14$ is $0.22\pm0.02$,\footnote{Although the value of $\alpha$ for $\Sigma_{10}/\mathrm{Mpc}^{-2}\ge12.26$ is consistent with that for $1.14\le\Sigma_{10}/\mathrm{Mpc}^{-2}<12.26$ within the error, the former is slightly larger than the latter. The use of a complete sample extending to lower masses of $\log(M_\mathrm{star}/M_{\odot})\sim9.0$, obtained from a deeper spectroscopic survey covering a large area, may help determine whether this slight difference in $\alpha$ is indeed significant.} while that of the lowest-density environments is $0.58\pm0.06$. Thus, the slope of the mass--size relation at the low-mass range is much closer to that of the high-mass range in the lowest-density environments than in the high-density environments. The pivot mass of the mass--size relation is $\log(M_p/M_{\odot})\approx10.3$, except for the highest-density bin which has a pivot mass of $\log(M_p/M_{\odot})=10.44\pm0.04$. The pivot mass of the lowest-density bin has a large error of 0.2 dex. This is because the difference in the slope of the mass--size relation between the low-mass range and high-mass range is very small in the lowest-density environments.

We find that the $\alpha$ of quiescent galaxies in the high-density environments of $\Sigma_{10}/\mathrm{Mpc}^{-2}\ge1.14$ ($0.22\pm0.02$) is similar to that of star-forming galaxies in the same environment ($0.21\pm0.01$). This finding suggests a connection between the two populations at the low-mass range of $\log(M_\mathrm{star}/M_{\odot})\lesssim10.0$ in dense environments (see Section \ref{sec:discuss}).
\\

\begin{deluxetable*}{rcccc}
\tabletypesize{\scriptsize}
\tablecaption{Best-fit Parameters of the Smoothly Broken Power-law Functions \label{tb:best}}
\tablehead{\colhead{} & \colhead{$\alpha$} & \colhead{$\beta$} & \colhead{$\log(M_p/M_{\odot})$} & \colhead{$r_p / \mathrm{kpc}$}
}
\startdata
All & $0.237\pm0.018$ & $0.771\pm0.009$ & $10.27\pm0.02$ & $1.67\pm0.05$\\
$\Sigma_{10}\ge12.26$ & $0.240\pm0.026$ & $0.816\pm0.032$ & $10.44\pm0.04$ & $2.09\pm0.11$\\
$1.14\leq\Sigma_{10}<12.26$ & $0.213\pm0.028$ & $0.745\pm0.014$ & $10.25\pm0.03$ & $1.67\pm0.07$\\
$0.25\leq\Sigma_{10}<1.14$ & $0.351\pm0.038$ & $0.803\pm0.017$ & $10.30\pm0.04$ & $1.66\pm0.11$\\
$0\leq\Sigma_{10}<0.25$ & $0.576\pm0.060$ & $0.778\pm0.042$ & $10.39\pm0.18$ & $1.95\pm0.57$\\
\hline
Star-forming galaxies with $\Sigma_{10}\ge1.14$ & $0.210\pm0.010$ & $0.617\pm0.040$ & $10.46\pm0.04$ & $3.52\pm0.12$\\
Star-forming galaxies with $\Sigma_{10}<1.14$ & $0.206\pm0.008$ & $0.573\pm0.039$ & $10.47\pm0.04$ & $3.60\pm0.12$
\enddata
\tablecomments{See Equation (\ref{eq:bp}) for the smoothly broken power-law function. The units of $\Sigma_{10}$ are Mpc$^{-2}$.
}
\end{deluxetable*}

\section{Discussion}\label{sec:discuss}
It is possible to infer the formation mechanisms of galaxies based on the mass--size relation. For example, \citet{vanDokkum2015} suggest that the growth of galaxies predominantly through star formation causes a slope of 0.3 (stellar mass growth dominates) in the mass--size relation,\footnote{Our study and previous studies focusing on low redshifts \citep{Mosleh2013,Lange2015,Kawinwanichakij2021} suggest a shallower slope of $\lesssim0.2$ for the mass--size relation of low-mass (star-forming or quiescent) galaxies. The slight dissimilarity in the slope compared with \citet{vanDokkum2015} could be attributed to the different mass ranges of the samples. \citet{vanDokkum2015} deduced the slope from high-redshift ($z\gtrsim1.5$) star-forming galaxy samples mainly with stellar masses of $\log(M_\mathrm{star}/M_{\odot})\gtrsim10.0$ (see their Figure 22). In contrast, the studies conducted at lower redshifts encompass galaxies with lower masses of $\log(M_\mathrm{star}/M_{\odot})\lesssim9.6$, thereby contributing to the difference.} whereas growth primarily via dry mergers creates a slope of 2 (size growth dominates; \citealt{Bernardi2011a,Oogi2013,Yoon2017}). In addition, \citet{Robertson2006} demonstrate that mergers of gas-rich galaxies generate merger remnants with a mass--size relation as steep as that of early type galaxies.

The slope of the low-mass part ($\log(M_\mathrm{star}/M_{\odot})\lesssim10.0$) of the mass--size relation for quiescent galaxies in high-density environments is found to be $\alpha=0.22\pm0.02$. This shallow slope may indicate that stellar mass growth has been more dominant in LQGs with $\log(M_\mathrm{star}/M_{\odot})\lesssim10.0$ in high-density environments than size growth, and their growth mechanism is more likely to be related to star formation rather than mergers \citep{vanDokkum2015}. Additionally, LQGs in high-density environments have a very similar mass--size relation slope as low-mass star-forming galaxies in the same environments ($\alpha=0.21\pm0.01$). Furthermore, LQGs in higher-density environments have lower S{\'e}rsic indices, suggesting that the structures of LQGs are more similar to disks in denser environments. These lines of evidence support the idea that LQGs in high-density environments, such as clusters and groups, were once star-forming galaxies, but have experienced quenching of star formation due to several environmental effects in dense environments, resulting in their transformation into quiescent galaxies, as suggested by \citet{Kawinwanichakij2021}. 

Our findings are consistent with previous studies \citep{Barazza2002,Pedraz2002,DeRijcke2003,Graham2003,Lisker2006,Boselli2008,Toloba2011}, which show that a substantial fraction of LQGs or dwarf elliptical galaxies in cluster environments retain rotation-supported systems and disk-like structures, implying that their progenitors are likely to be star-forming disk galaxies. Similarly, \citet{Chen2022} show that low-mass poststarburst galaxies with $\log(M_\mathrm{star}/M_{\odot})\lesssim10.0$ residing in groups exhibit more disk-like structures and larger sizes for a given mass compared to their counterparts in field environments, suggesting that their origin is related to environment-driven fast quenching.

Although the slopes of the mass--size relations for LQGs in dense environments and star-forming galaxies are very similar to each other, LQGs in high-density environments exhibit smaller sizes (or a lower normalization in the relation) than star-forming galaxies in the same environments, for a given mass. This can be straightforwardly elucidated within the context of the cosmic evolution of galaxy sizes, in which star-forming galaxies were smaller in the past \citep{Dahlen2007,Williams2010,Mosleh2011,Morishita2014,Ribeiro2016,Allen2017,Paulino2017,Kawinwanichakij2021}.

The evolution of galaxy sizes is quantified by the parameterization $R_e\propto(1+z)^l$, with reported values of $l$ ranging approximately from $-0.5$ to $-1.3$ for star-forming galaxies across different studies. If we choose $l=-1.1$ \citep{Dahlen2007,Williams2010,Mosleh2011}, the median size difference of 0.2 dex between LQGs in the highest-density environments ($\Sigma_{10}\ge12.26$) and low-mass star-forming galaxies implies that star-forming galaxies ceased to grow in size at $z\sim0.6$ ($\sim6$ Gyr ago)\footnote{Alternatively, $z\sim1.0$ ($\sim8$ Gyr ago), if $l$ is $-0.75$. } and transitioned into LQGs in the local Universe over timescales of a few gigayears through several environmental effects in dense environments \citep{Boselli2006}. Then, star-forming galaxies that experienced the same processes earlier or later than $z\sim0.6$ would exhibit slightly different sizes depending on the transformation epochs. As a result, they could contribute to the scatter in the mass--size relation. This is also consistent with the fact that LQGs with bluer $u-r$ colors (younger stellar populations) have larger sizes (Appendix \ref{Appendix_B}).

In the same context, low-mass galaxies at high redshifts of $z\gtrsim2$ ($\sim10$ Gyr ago) have very compact sizes. However, it is unlikely that they are quenched by environmental effects in the same manner as in the low-redshift Universe and transition into LQGs in high-density environments. This is because it is hard to quench low-mass galaxies by environmental effects at such high redshifts, as the environmental quenching in dense environments is only efficient at lower redshifts of $z\lesssim1.5$ \citep{Peng2010,Lee2015,Kawinwanichakij2017}, when cluster-like large-scale structures have matured. Moreover, since the merger rate and star formation rate of galaxies are high at such high redshifts \citep{Behroozi2013,Madau2014,Rodriguez2015,Yoon2017,Ferreira2020}, a large fraction of low-mass galaxies at high redshifts are likely to experience galaxy mergers and substantial growth, transitioning into more massive galaxies. For example, using the galaxy formation simulation of \citet{Guo2011} based on the Millennium-II simulation \citep{Boylan2009}, we find that roughly $80\%$ of low-mass galaxies with $9.4\le\log(M_\mathrm{star}/M_{\odot})<9.8$ at the high redshift of $z\sim2.5$ evolve into galaxies with $\log(M_\mathrm{star}/M_{\odot})>10.2$ in the local Universe.

By contrast, star-forming galaxies in dense environments at $z\lesssim0.2$ (equivalent to a lookback time of $\lesssim2.4$ Gyr) may not have had enough time to be quenched in the local Universe. We note that the star formation of satellite galaxies can be unaffected for $2$--$4$ Gyr after their infall into clusters or groups, according to \citet{Wetzel2013}.

Additionally, there is the possibility that environmental effects in dense environments can reduce galaxy sizes. For example, the stripping or removal of less strongly bound gas in the outer regions of galaxies can cause the luminosity of disks to fade out (hence, the central regions of galaxies become more dominant), leading to a reduction of the sizes of galaxies \citep{Kuchner2017}. If this effect is indeed capable of decreasing galaxy sizes by $\gtrsim20\%$, our result can also be in agreement with the slow size evolution ($l\gtrsim-0.4$ in $R_e\propto(1+z)^l$) that low-mass galaxies may have experienced (e.g., \citealt{Kawinwanichakij2021}).

LQGs in the lowest-density environments of $\Sigma_{10}<0.25$ not only have very small sizes, but also exhibit a steep slope ($\alpha=0.58\pm0.06$) in their mass--size relation. This requires another mechanism apart from environmental quenching that is suitable for explaining the origin of the LQGs in dense environments. The simulation of \citet{Robertson2006} shows that remnants of a single gas-rich major merger form a mass--size relation with a normalization lower than that of progenitor disk galaxies. This is because gas-rich mergers are able to generate high gas densities in the central regions of remnants \citep{Hernquist1989,Barnes1991,Barnes1996}, leading to a correspondingly high star formation rate.  As a result, this causes high stellar mass densities in those central regions, and consequently, smaller galaxy sizes for a given mass. Moreover, this effect is more prominent in lower-mass galaxies, which leads to the formation of a steep slope in the mass--size relation. This is because a larger fraction of the gas is heated and remains in the halo during the merger process of more massive galaxies, while cold gas is more abundant in lower-mass galaxies \citep{Bell2000,Masters2012}. Specifically, the simulation of \citet{Robertson2006} demonstrates that the remnants of a single gas-rich major merger\footnote{The gas fractions of mergers are 0.4 and 0.8 in the simulation.} form a mass--size relation with a slope of $0.57$, which is exactly consistent with that of LQGs in the lowest-density environments, $0.58\pm0.06$. Similarly, the simulation of  \citet{Covington2011} demonstrates that it is essential to include gas-rich major mergers and the decreasing gas abundance as a function of the mass of progenitors, in order to turn the shallow mass--size relation of disk galaxies into a steep mass--size relation with a lower normalization (a single merger is sufficient). Thus, according to the simulations, gas-rich mergers are the possible origin of the mass--size relation with a steep slope and a low normalization that LQGs in the lowest-density environments exhibit.

\begin{figure*}
\includegraphics[scale=0.25]{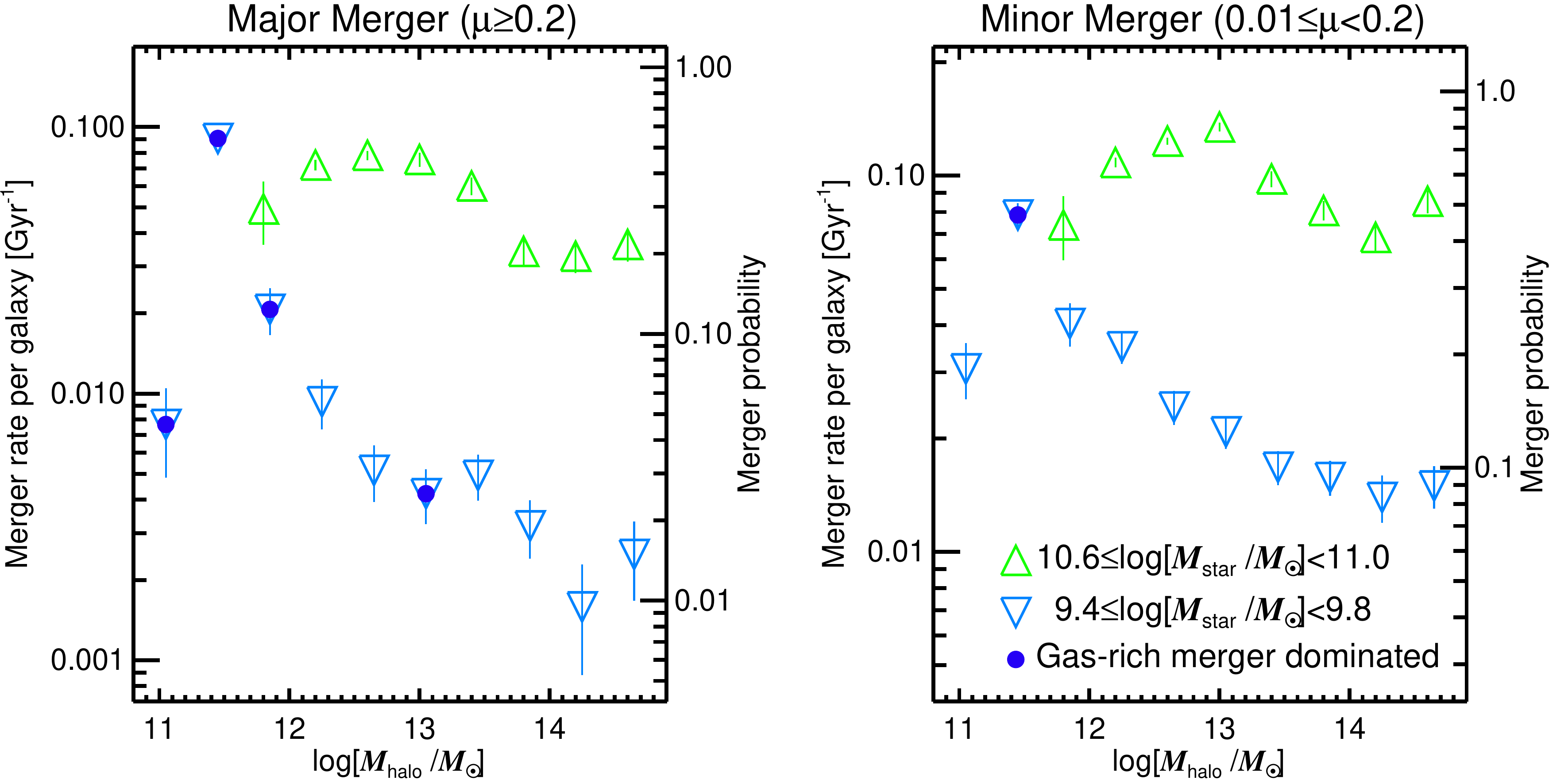}
\includegraphics[scale=0.25]{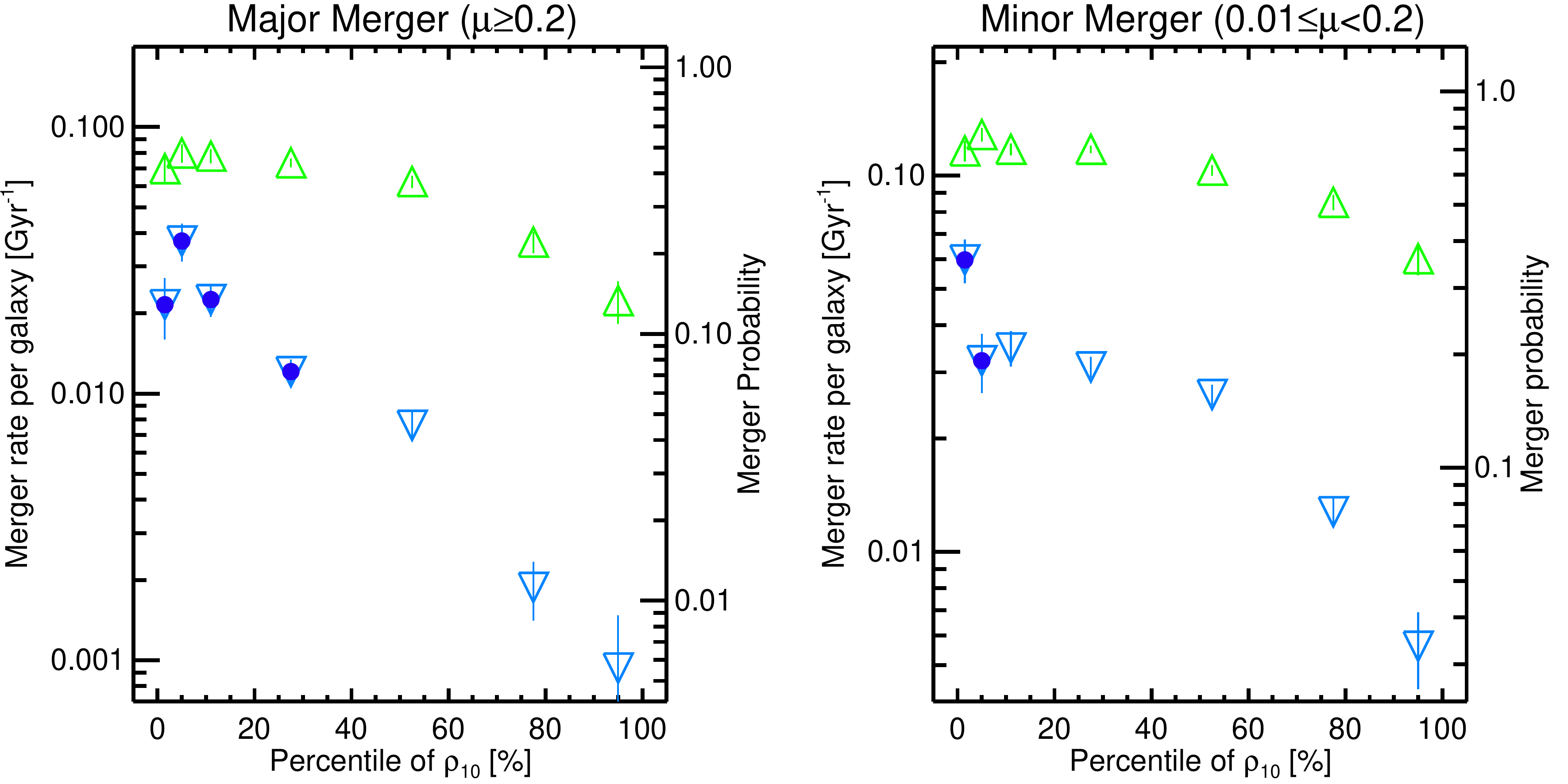}
\centering
\caption{Merger rate per galaxy (left $y$-axes) and merger probability (right $y$-axes) for simulated galaxies as a function of halo mass $M_\mathrm{halo}$ (upper panels) and local number density of galaxies $\rho_{10}$ (lower panels). The results for major mergers ($\mu\ge0.2$) are shown in the left panels, while the results for minor mergers ($0.01\le\mu<0.2$) are displayed in the right panels. The target galaxies used in this figure are quiescent galaxies at $z=0$ from the galaxy formation simulation of \citet{Guo2011} based on the Millennium-II simulation \citep{Boylan2009}. The mergers that occur by $z<0.63$ are counted to calculate the merger rates of the target galaxies. The data points are differentiated by shape (and color) to indicate the different stellar mass bins. The bins in which more than $75\%$ of mergers are gas-rich ones are marked by the blue filled circles.
\label{fig:merger}}
\end{figure*}

However, one thing that should be addressed is whether mergers can occur frequently in low-density environments. In order to examine this question, we analyze the merger rates of simulated galaxies in various environments, using the cosmological galaxy formation simulation of \citet{Guo2011} based on the Millennium-II simulation \citep{Boylan2009}. The targets are quiescent galaxies at $z=0$ with specific star formation rates (sSFRs) of less than $10^{-11}$yr$^{-1}$.\footnote{In the simulation, the star-forming main sequence is at sSFR$>10^{-10.5}$yr$^{-1}$. Thus, the criterion sSFR$<10^{-11}$yr$^{-1}$ effectively distinguishes quiescent galaxies from star-forming ones.} The mergers that occur at $z<0.63$  (roughly over a period of $6$ Gyr) are counted to compute the merger rates of the target galaxies. Thus, we only look into mergers in the low-redshift Universe.  Major mergers and minor mergers are defined by $\mu\ge0.2$ and $0.01\le\mu<0.2$, respectively, where $\mu$ is the stellar mass ratio between the secondary galaxy and the primary galaxy. Here, we define gas-rich mergers as mergers with $f_\mathrm{gas}\ge0.8$, where $f_\mathrm{gas}$ is the ratio between the mass of cold gas and the total mass of stars and cold gas, for merger progenitors.

The upper panels of Figure \ref{fig:merger} display the merger rate per galaxy ($R_m$) and merger probability\footnote{The merger probability is defined as the total number of mergers experienced by target galaxies at $z<0.63$ divided by the number of target galaxies.} for simulated galaxies as a function of halo mass ($M_\mathrm{halo}$). The halo mass represents the virial mass of the group in which the target galaxies reside at $z=0$. The figure shows that HQGs with $10.6\le\log(M_\mathrm{star}/M_{\odot})<11.0$ have the highest major merger rate of $R_m=0.075$ Gyr$^{-1}$ in halos with $12.0<\log(M_\mathrm{halo}/M_{\odot})<13.2$, which corresponds to small group-scale environments. We note that HQGs in the halo mass range of $12.0<\log(M_\mathrm{halo}/M_{\odot})<13.2$ experience at least one merger (either major or minor) by $z<0.63$. In lower-mass halos with $\log(M_\mathrm{halo}/M_{\odot})<12.0$, the major merger rate per galaxy of HQGs is lower ($R_m=0.049$ Gyr$^{-1}$). Similarly, HQGs experience a small number of major mergers ($R_m=0.034$ Gyr$^{-1}$) in rich-group and cluster environments with $\log(M_\mathrm{halo}/M_{\odot})>13.6$. The almost same trend is also found in the case of the minor merger rate of HQGs ($R_m=0.12$ Gyr$^{-1}$ at $12.0<\log(M_\mathrm{halo}/M_{\odot})<13.2$ and $R_m=0.078$ Gyr$^{-1}$ at $\log(M_\mathrm{halo}/M_{\odot})>13.6$), except that the minor merger rate is higher than the major merger rate.

This trend is the consequence of the balance between the dynamical merger timescale and the galaxy abundance in the halo of a certain mass (see \citealt{Hopkins2008}). The merger timescale for a galaxy of a given stellar mass is long in massive host halos, owing to inefficient dynamical friction (in very massive halos, galaxies move too fast for their stellar masses for mergers to occur). Conversely, in a low-mass host halo, the probability that the halo contains numerous companion galaxies around a galaxy of a given stellar mass declines. As a result, the merger rate of galaxies has a peak at which the two effects are in balance, and as shown above, HQGs with $10.6\le\log(M_\mathrm{star}/M_{\odot})<11.0$ in the local Universe have a peak merger rate in small group-scale environments.\footnote{At high redshifts of $z\gtrsim2$, halos destined to evolve into galaxy clusters in the local Universe possess halo masses that are low enough ($\log(M_\mathrm{halo}/M_{\odot})<13.5$) for galaxy mergers to occur at high rates.}

The peak of merger rates is expected to shift toward lower-mass halos as the stellar masses of the target galaxies decrease. This is because the merger timescale is short enough only in low-mass halos for the mergers of low-mass galaxies to occur frequently, while low-mass halos can still accommodate many companion galaxies around a low-mass galaxy. The result for LQGs shown in the upper panels of Figure \ref{fig:merger} exactly illustrates this trend. The peak of the merger rate for LQGs with $9.4\le\log(M_\mathrm{star}/M_{\odot})<9.8$ is at $\log(M_\mathrm{halo}/M_{\odot})\approx11.5$ for both major and minor mergers. At this peak, the major merger rate is $R_m=0.091$ Gyr$^{-1}$, while the minor merger rate is $R_m=0.079$ Gyr$^{-1}$. Thus, the merger rates of LQGs at $\log(M_\mathrm{halo}/M_{\odot})\approx11.5$ are comparable to, or even higher than, those of HQGs. In other words, LQGs at $\log(M_\mathrm{halo}/M_{\odot})\approx11.5$ undergo at least one merger (either major or minor) by $z<0.63$. However, the merger rates decline very rapidly as the host halo mass of LQGs decreases or increases from $\log(M_\mathrm{halo}/M_{\odot})\sim11.5$. In cluster environments with $\log(M_\mathrm{halo}/M_{\odot})>14.0$, the merger rates of LQGs are extremely low ($R_m=0.002$ Gyr$^{-1}$ for major mergers and $R_m=0.014$ Gyr$^{-1}$ for minor mergers), so that they are about an order of magnitude lower than those of HQGs.

We calculate the local galaxy number density of simulated galaxies within the three-dimensional distance to the tenth-nearest galaxy ($\rho_{10}$) and examine the merger rate per galaxy as a function of $\rho_{10}$. The local number density $\rho_{10}$ is computed using galaxies with  $\log(M_\mathrm{star}/M_{\odot})\ge9.4$ as in Section \ref{sec:env}. The lower panels of Figure \ref{fig:merger} display the results. Since the local number density of galaxies has a good correlation with the host halo mass (the Spearman's rank correlation coefficient is 0.79), the results for $\rho_{10}$ are consistent with those for $M_\mathrm{halo}$. For instance, the peak of the merger rate of LQGs is found at the low $\rho_{10}$ percentile of $<10\%$, and the merger rate declines rapidly as $\rho_{10}$ increases for both major and minor mergers. Specifically, the major and minor merger rates of LQGs at the low $\rho_{10}$ percentile of $<7\%$ are $33$ times and $8$ times higher, respectively, than those at the high $\rho_{10}$ percentile of $>90\%$. However, the sharp peaks in the major and minor merger rates detected in the LQGs at $\log(M_\mathrm{halo}/M_{\odot})\approx11.5$ are somewhat diminished in the result of $\rho_{10}$. This is because the low $\rho_{10}$ range with the percentile of $<10\%$ covers up to $\log(M_\mathrm{halo}/M_{\odot})\sim12.7$ due to the scatter in the relation between $M_\mathrm{halo}$ and $\rho_{10}$.

We also examine the contribution of gas-rich mergers to the merger rate (see the blue filled circles in Figure \ref{fig:merger}).  The contribution of gas-rich mergers is higher in lower-mass galaxies due to a well-known trend that lower-mass galaxies are observed to have higher gas fractions \citep{Masters2012}, which is also confirmed in the simulation data used here. For HQGs, over $\sim80\%$ of major or minor mergers are dry mergers, regardless of the host halo mass. By contrast, most mergers that LQGs have experienced over the last $6$ Gyr are gas-rich mergers. Specifically, $90\%$ of the major mergers that LQGs in halos with $\log(M_\mathrm{halo}/M_{\odot})<12.0$ undergo are gas-rich mergers. In more massive halos with $\log(M_\mathrm{halo}/M_{\odot})>12.0$, this fraction drops to $60\%$. Although the fraction of gas-rich mergers is relatively low in the case of minor mergers that LQGs experience, it is still higher than that of HQGs in any halo mass bin, particularly in low-mass halos with $\log(M_\mathrm{halo}/M_{\odot})<11.8$, where the fraction of gas-rich mergers is $75\%$. This trend for the fraction of gas-rich mergers is also similarly observed in the result for $\rho_{10}$, so that the merger rates of LQGs in low-density environments are dominated by gas-rich mergers. We note that the star formation rates induced by gas-rich mergers are quickly suppressed by supernova feedback in the case of low-mass galaxies \citep{Guo2011}, leading to quiescent remnants at $z=0$.

With the aid of the simulation data, we find that LQGs in the lowest-density environments are over 10 times more likely to have experienced recent gas-rich mergers than their counterparts in cluster-like dense environments ($\gtrsim30$ times in the case of gas-rich major mergers). Thus, the analysis of merger histories in the cosmological galaxy formation simulation, in combination with the results of the simulations of \citet{Robertson2006} and \citet{Covington2011} described above, suggests that the steep slope and the low normalization of the mass--size relation observed in LQGs in the lowest-density environments may originate from recent gas-rich mergers (with the trend for the gas fraction to decrease in more massive galaxies).

The simulation also demonstrates that cluster-like dense environments are unsuitable places for mergers to occur for LQGs. Therefore, when combined with our observational result, this indicates that the influence of galaxy mergers is minimal while the environmental effects of dense environments are maximal in LQGs in the highest-density environments.

LQGs in group-scale intermediate environments experience nonnegligible (gas-rich) mergers. Moreover, they can be influenced by moderately efficient environmental effects \citep{Fujita2004,Boselli2006}. Hence, the gradual shift in the slope and normalization of the mass--size relation from LQGs in low-density to high-density environments can be attributed to the gradual changes in the influences of the two distinct factors: mergers and environmental effects.

Finally, the simulation shows that in the low-redshift Universe, mergers can occur even in environments with very low local densities and affect the evolution of quiescent galaxies there. This is further supported by a previous study based on a cosmological simulation \citep{Niemi2010}, as well as by observational studies showing that traces of recent galaxy mergers, such as tidal features, are frequently discovered around early type galaxies in isolated environments \citep{Reduzzi1996,Colbert2001,Kuntschner2002,Reda2004,Reda2005,Hernandez2008}.

\section{Summary}\label{sec:summary}
We investigate the mass--size relation of quiescent galaxies across different environments, using a large data set from SDSS. Our sample comprises 13,667 quiescent galaxies with $\log(M_\mathrm{star}/M_{\odot})\ge9.4$ and $0.01<z<0.04$. Quiescent galaxies are classified using the $u-r$ versus $r-z$ color--color diagram with the aid of EW(H$\alpha$) information. We define the environment of a galaxy as the surface number density of galaxies within a rest-frame velocity range of $\pm1000$ km s$^{-1}$ and the distance to the tenth-nearest galaxy ($\Sigma_{10}$). Alternatively, the use of the surface stellar mass density of the ten nearest neighbor galaxies ($\Sigma_{M}$) results in nearly identical outcomes. A smoothly broken power-law function is used to examine the slope of the mass--size relation for quiescent galaxies and its variation across different mass ranges. Our main results are as follows.
\begin{enumerate}
\item The mass--size relation of LQGs with $\log(M_\mathrm{star}/M_{\odot})\lesssim10.0$ depends on environment, in such a way that LQGs are smaller in lower-density environments for a given mass. Specifically, the mean $R_e$ for LQGs with $\log(M_\mathrm{star}/M_{\odot})\lesssim9.8$ in the lowest-density environments is $60\%$ of that for their counterparts in the highest-density environments.

\item The slope of the mass--size relation for LQGs with $\log(M_\mathrm{star}/M_{\odot})\lesssim10.0$ in high-density environments ($0.22\pm0.02$) is significantly shallower than that of their counterparts in isolated environments ($0.58\pm0.06$), but very similar to that of low-mass star-forming galaxies in the same environments ($0.21\pm0.02$). 

\item LQGs in high-density environments tend to have structures that are more disk-like (lower S{\'e}rsic indices) than their counterparts in low-density environments. 

\item HQGs with $\log(M_\mathrm{star}/M_{\odot})\gtrsim10.5$ exhibit a nearly identical mass--size relation across all environments, with a slope of $\approx0.77$.
\end{enumerate}

Our observational results, combined with the analysis of merger rates for simulated galaxies in a cosmological galaxy formation simulation, suggest implications for the formation of LQGs as follows. 
\begin{enumerate}
\item LQGs in high-density environments were once star-forming galaxies, but have undergone star formation quenching due to environmental effects in dense environments, ultimately transforming into quiescent galaxies without significant merger events.

\item The steep slope and low normalization of the mass--size relation of LQGs in the lowest-density environments may stem from recent gas-rich mergers, which occur over $10$--$30$ times more frequently in the progenitors of LQGs in the lowest-density environments than in their counterparts in high-density environments at low redshifts.
\end{enumerate}

\begin{acknowledgments}
This research was supported by the Korea Astronomy and Space Science Institute under the R\&D program (Project No. 2023-1-830-00), supervised by the Ministry of Science and ICT.
The Millennium-II Simulation databases used in this paper and the web application providing online access to them were constructed as part of the activities of the German Astrophysical Virtual Observatory (GAVO).\\
\end{acknowledgments}

\appendix

\begin{figure*}
\includegraphics[scale=0.30]{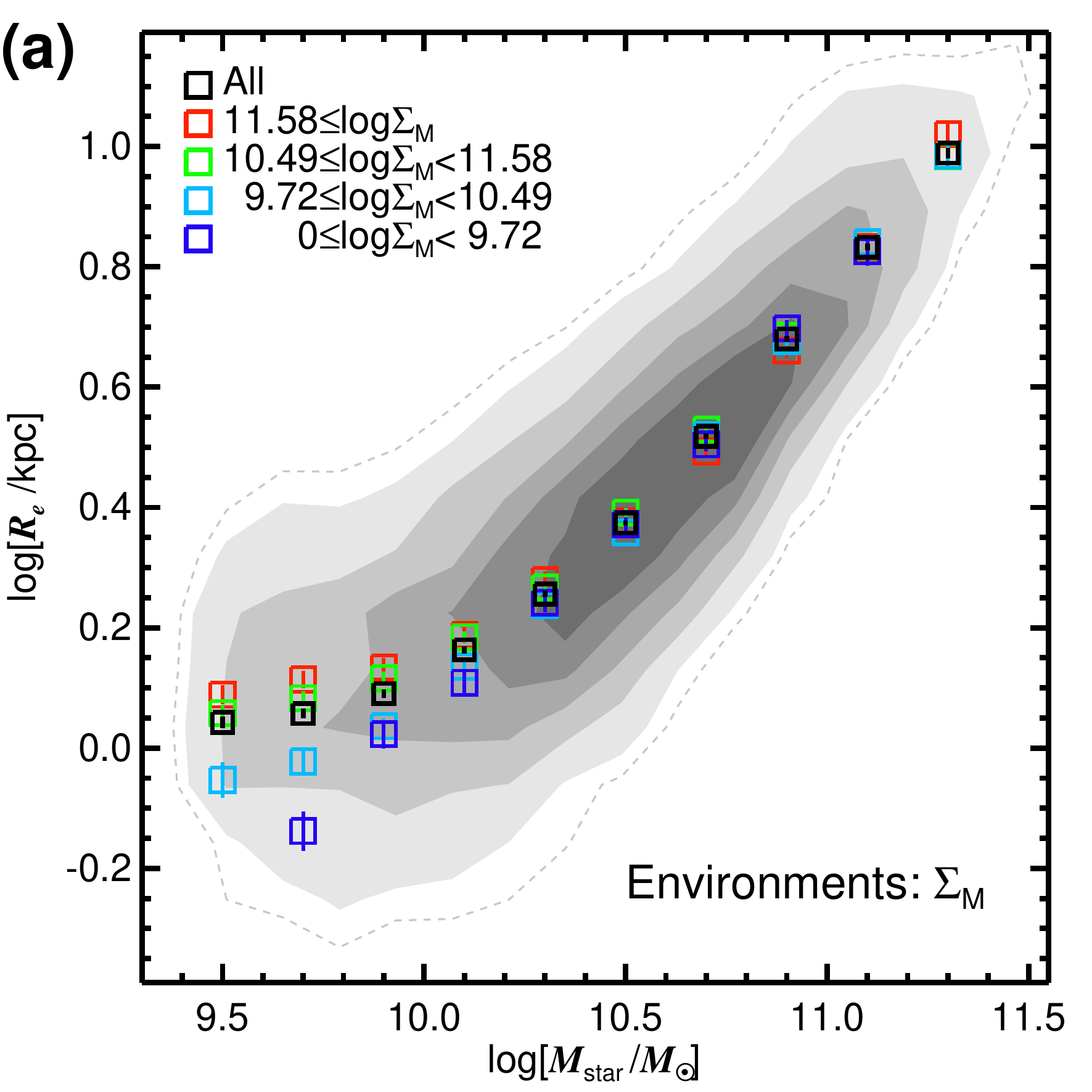}\includegraphics[scale=0.30]{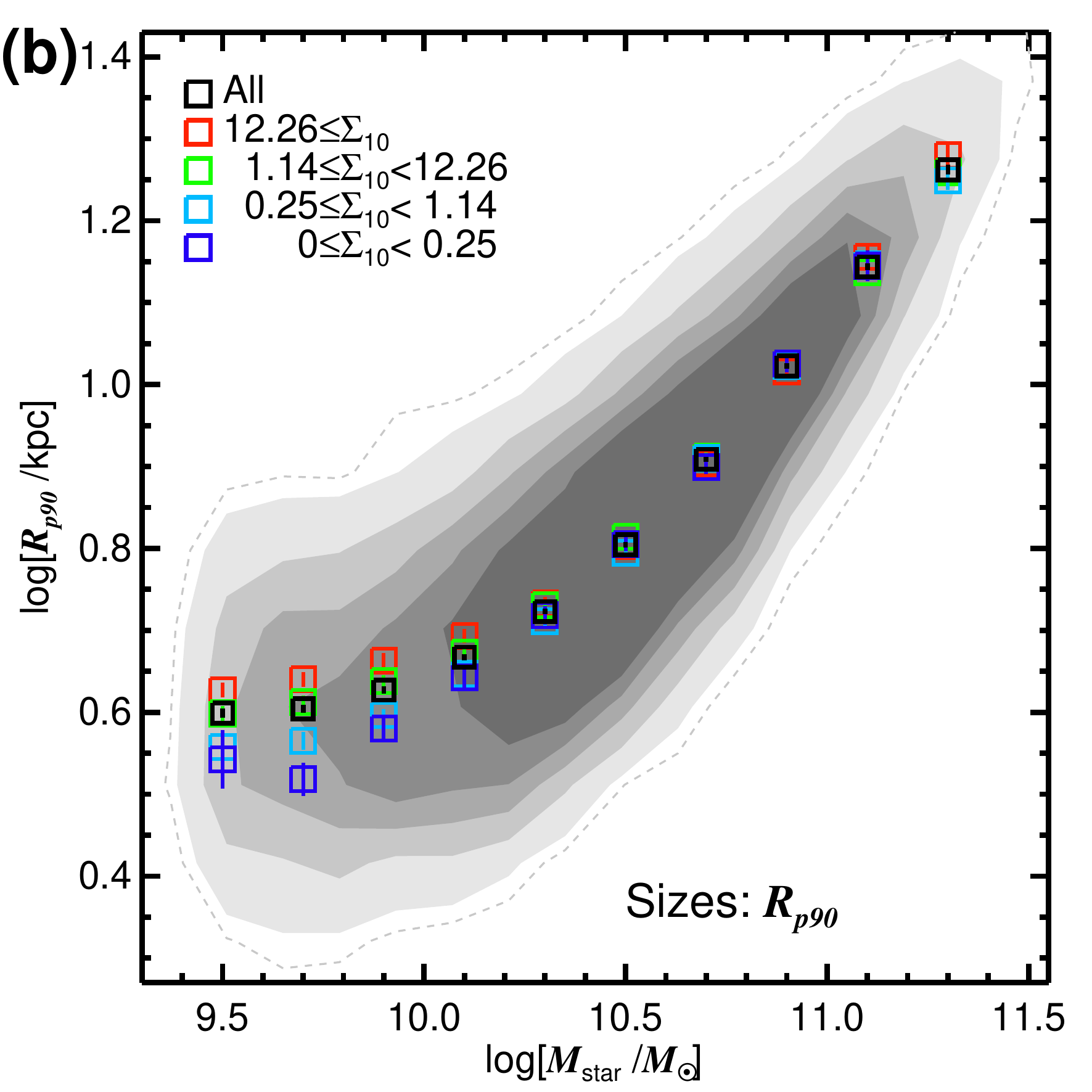}  
\includegraphics[scale=0.30]{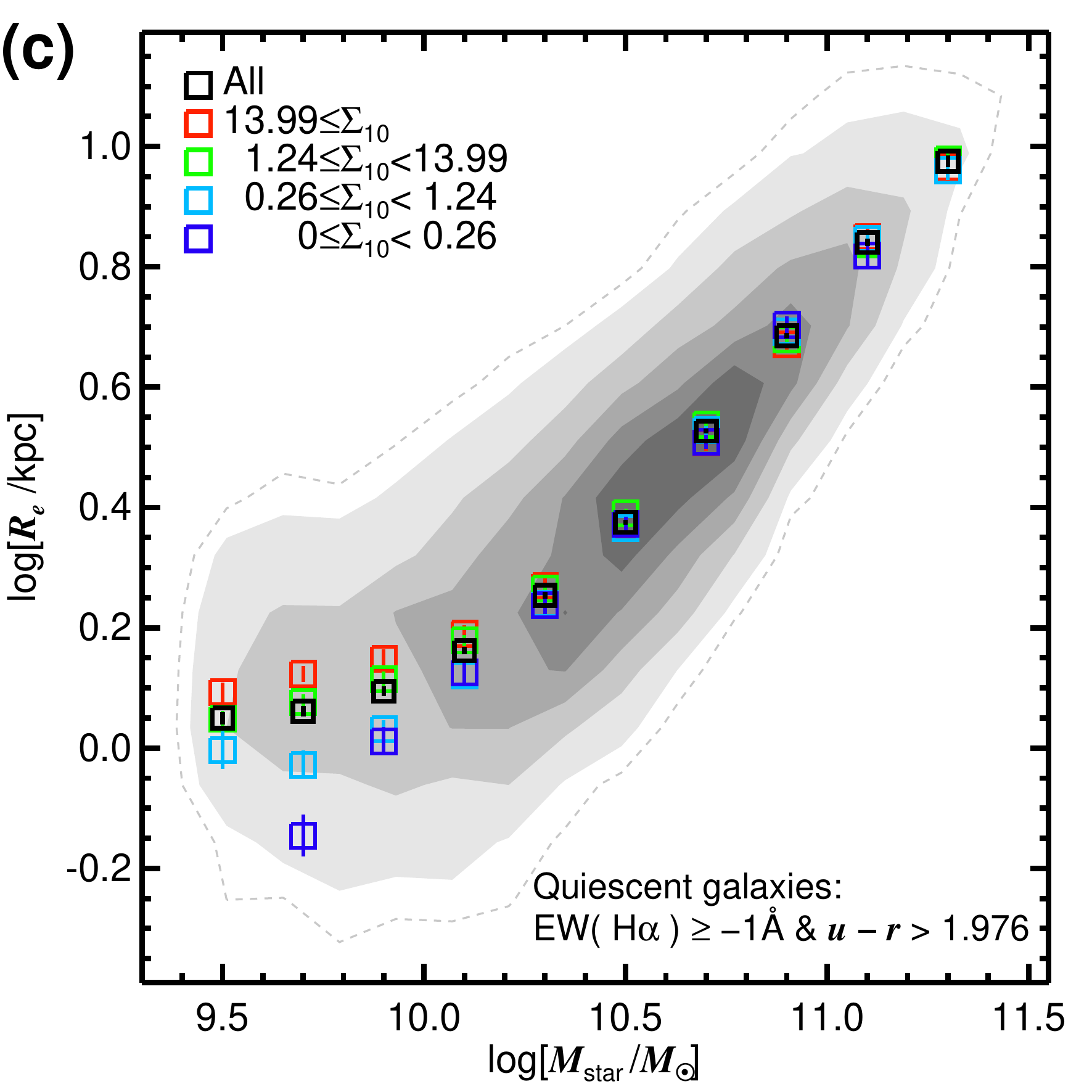}\includegraphics[scale=0.30]{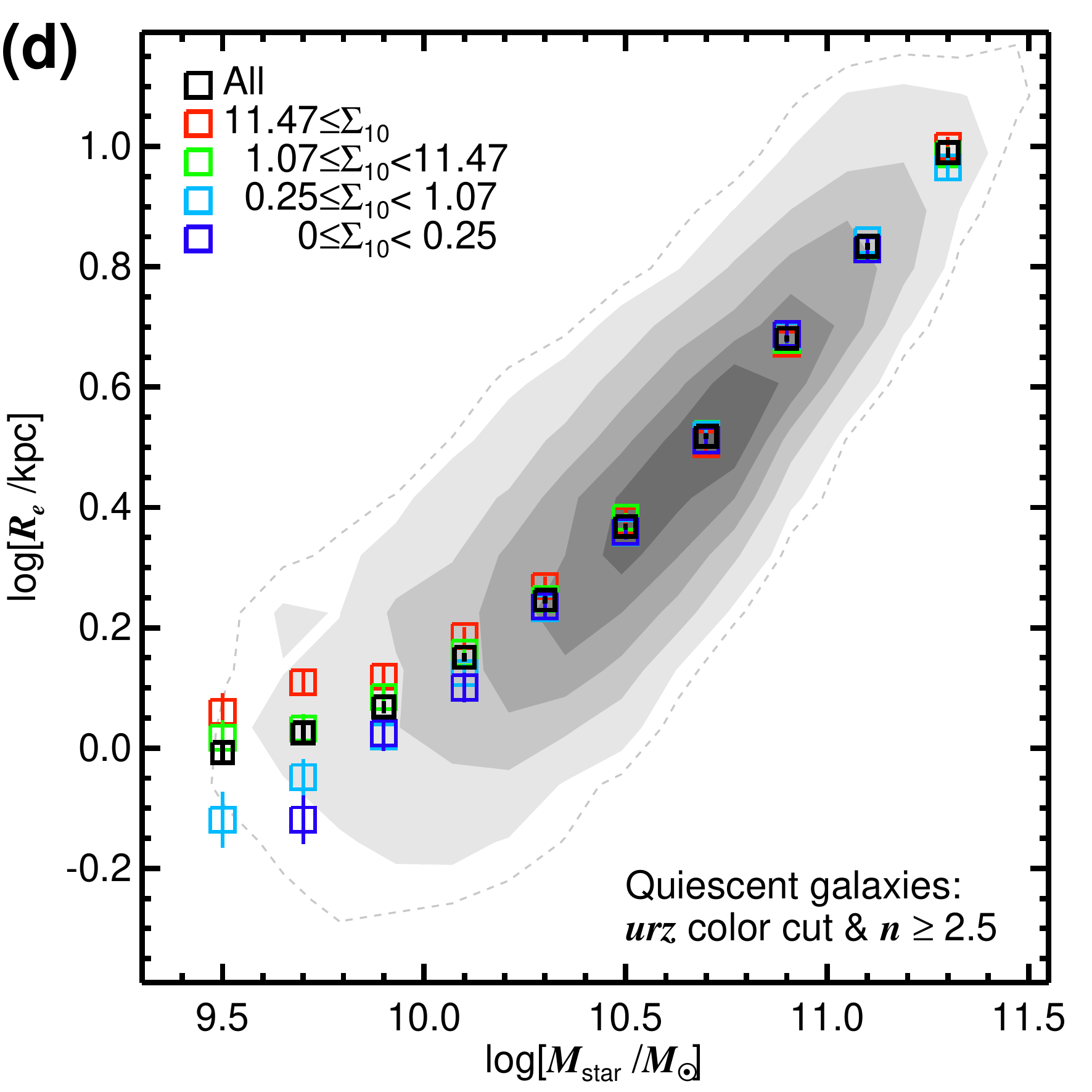} 
\centering
\caption{Mass--size relations of quiescent galaxies in various environments, when using alternative definitions for quiescent galaxies, size, and environment. Panel (a): the case when $\Sigma_{M}$ (Equation (\ref{eq:denm})) is used as the definition of environment instead of $\Sigma_{10}$. The units of $\Sigma_{M}$ are $M_{\odot}$ Mpc$^{-2}$. Panel (b): the case when the $r$- band Petrosian $90\%$ light radius ($R_{p90}$) is used instead of $R_e$. Panel (c): the case when the main criterion to select quiescent galaxies is EW(H$\alpha$)$\ge-1\mathrm{\AA}$. Panel (d): the case when quiescent galaxies are selected using the combination of a S{\'e}rsic index cut of $n\ge2.5$ and the color cut of Equations (\ref{eq:cut1}) and (\ref{eq:cut2}). The four environment bins in each panel are set to correspond to percentile ranges of 0 -- 10\%, 10 -- 40\%, 40 -- 80\%, and 80 -- 100\%, respectively, as in the main text. The other details in each panel are the same as those in Figure \ref{fig:ms}.
\label{fig:ms_a}}
\end{figure*} 

\begin{figure}
\includegraphics[width=\linewidth]{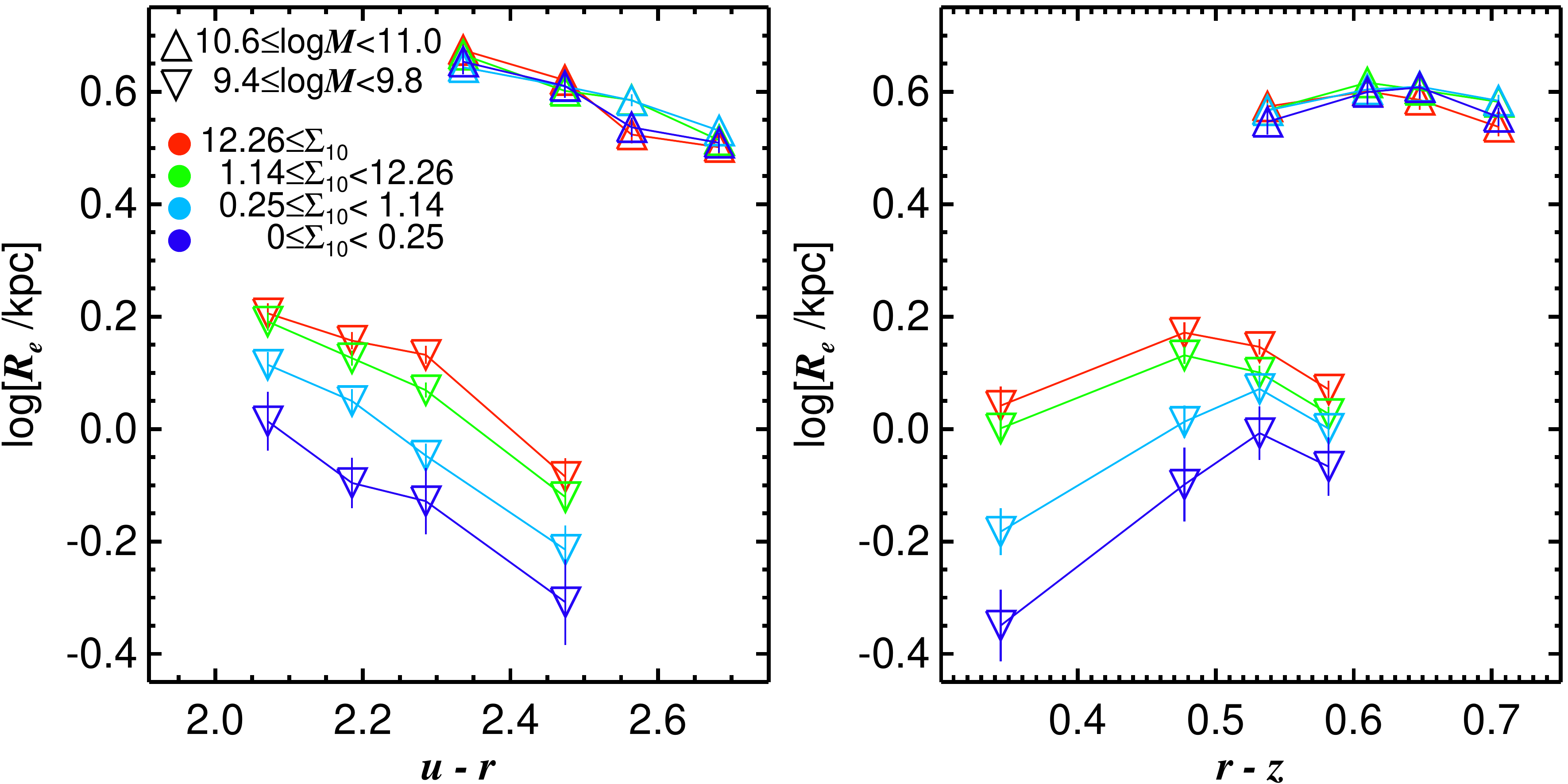}
\caption{Correlation between the colors and sizes of our quiescent galaxies in the different environments (the left panel is for $u-r$ and the right panel is for $r-z$). Each data point (the triangle) represents the mean $\log R_e$ of the color bin, after excluding outliers beyond $5\sigma$. The error bars denote the standard error of the mean. The data points are differentiated by shape to indicate the different stellar mass bins and by colors to represent the different environment bins. The four color bins correspond to percentile ranges of 0 -- 25\%, 25 -- 50\%, 50 -- 75\%, and 75 -- 100\%, respectively, for the colors of LQGs and HQGs. The units of $\Sigma_{10}$ in the legend are Mpc$^{-2}$.
\label{fig:sizecolor}}
\end{figure} 

\section{Mass--Size Relations from Alternative Definitions for Sample, Size, and Environment}\label{Appendix_A}
In Figure \ref{fig:ms_a}, we present the mass--size relations of quiescent galaxies in various environments, which are based on alternative definitions for quiescent galaxies, size, and environment. Panel (a) of Figure \ref{fig:ms_a} displays the case when  $\Sigma_{M}$ (Equation (\ref{eq:denm})) is used as the definition of galaxy environment instead of $\Sigma_{10}$. Panel (b) shows the case when the $r$-band Petrosian $90\%$ light radius (the radius containing 90\% of the Petrosian flux; $R_{p90}$) is used instead of the half-light radius from S{\'e}rsic model fits. We note that the seeing effect is not corrected in $R_{p90}$. Panel (c) illustrates the case when selecting quiescent galaxies based on the criterion of EW(H$\alpha$)$\ge-1\mathrm{\AA}$. We find that some galaxies selected by this criterion exhibit blue outer regions, likely due to the fact that the EW(H$\alpha$) values are derived from the central regions ($<1.5\arcsec$) of the galaxies. To prevent misclassification of such galaxies as quiescent, we implement an additional requirement of $u-r > 1.976$ (Equation \ref{eq:cut1}), which helps to avoid the identification of galaxies with quiescent centers but blue, star-forming outer regions as quiescent galaxies. Panel (d) shows the result when quiescent galaxies are selected using the combination of a S{\'e}rsic index cut of $n\ge2.5$ and the color cut of Equations (\ref{eq:cut1}) and (\ref{eq:cut2}). The criterion of $n\ge2.5$ has been widely used to select spheroid-dominated galaxies \citep{Shen2003,Huertas-Company2013,Cebrian2014,Lange2015,Kuchner2017}. However, the simple S{\'e}rsic index cut may not be a reliable separator, particularly for low-mass galaxies with $\log(M_\mathrm{star}/M_{\odot})\lesssim10.0$, since the S{\'e}rsic index depends on stellar mass \citep{GG2003,Lange2015,Kuchner2017}.

Even though the alternative definitions for the quiescent sample, size, and environment are used, the results are still consistent with those presented in the main text. For example, the mass--size relation of HQGs with $\log(M_\mathrm{star}/M_{\odot})\gtrsim10.5$ is uniform across environments in all cases. In addition, the sizes of LQGs with $9.4\le\log(M_\mathrm{star}/M_{\odot})<9.8$ in the lowest-density environments are, on average, 0.22 -- 0.25 dex smaller than those in the highest-density environments in all the cases, except for the case when $R_{p90}$ is used, where the average size difference between the two extreme environments is reduced to 0.11 dex. The K-S tests of the size distributions of LQGs in the highest- and lowest-density bins yield $p<2\times10^{-9}$ in all cases (including the case when $R_{p90}$ is used), indicating that the size difference between LQGs in the two extreme environment bins is very significant even with the alternative definitions.
\\

\section{Correlation between Color and Size in Different Environments}\label{Appendix_B}

There is a possible correlation between the colors and sizes of quiescent galaxies. For example, \citet{Morishita2017} showed a $U-V$ color--size correlation in low-mass passive cluster galaxies with $\log(M_\mathrm{star}/M_{\odot})<9.8$, such that bluer galaxies have larger radii. Thus, we examine the correlation between the colors and sizes of our quiescent galaxies in the different environments, and test the influence of this correlation and color difference between the environments on our main results.

Figure \ref{fig:sizecolor} displays the correlation between the colors ($u-r$ and $r-z$) and sizes of our quiescent galaxies in the different environments. The figure shows that there is a correlation between $u-r$ color and $\log R_e$, where bluer quiescent galaxies tend to be larger than their redder counterparts by about 0.3 dex for LQGs with $9.4\le\log(M_\mathrm{star}/M_{\odot})<9.8$ and 0.15 dex for HQGs with $10.6\le\log(M_\mathrm{star}/M_{\odot})<11.0$. HQGs follow a nearly identical $u-r$ color--size correlation regardless of their environment. By contrast, the sizes of LQGs are consistently smaller in lower-density environments across all the $u-r$ color bins. 

While the $r-z$ color--size correlation is very weak for HQGs, the correlation is substantial for LQGs in low-density environments. However, the trend of the $r-z$ color--size correlation for LQGs in low-density environments is opposite to the $u-r$ color--size correlation, in the sense that  bluer quiescent galaxies tend to be smaller than their redder counterparts by $\sim0.13$ -- $0.34$ dex (with a larger difference in lower-density environments). As in the $u-r$ case, the sizes of LQGs are also smaller in lower-density environments, for a given color in all the $r-z$ color bins. 
 
Therefore, the slight $u-r$ and $r-z$ color differences between environments reported in Section \ref{sec:env} (maximum difference of 0.04) are only capable of producing a minor effect on the size difference between the different environments. By matching the color distributions of the different environment bins using a resampling method, we discover that the color differences between the different environments can account for only $\sim8$ -- $16\%$ of the observed size difference in LQGs at most.
\\


\clearpage

\begin{thebibliography}{}
\bibitem[Allen et al.(2017)]{Allen2017} Allen, R.~J., Kacprzak, G.~G., Glazebrook, K., et al.\ 2017, \apjl, 834, L11. doi:10.3847/2041-8213/834/2/L11
\bibitem[Barazza et al.(2002)]{Barazza2002} Barazza, F.~D., Binggeli, B., \& Jerjen, H.\ 2002, \aap, 391, 823. doi:10.1051/0004-6361:20020875
\bibitem[Barnes \& Hernquist(1991)]{Barnes1991} Barnes, J.~E. \& Hernquist, L.~E.\ 1991, \apjl, 370, L65. doi:10.1086/185978
\bibitem[Barnes \& Hernquist(1996)]{Barnes1996} Barnes, J.~E. \& Hernquist, L.\ 1996, \apj, 471, 115. doi:10.1086/177957
\bibitem[Behroozi et al.(2013)]{Behroozi2013} Behroozi, P.~S., Wechsler, R.~H., \& Conroy, C.\ 2013, \apj, 770, 57. doi:10.1088/0004-637X/770/1/57
\bibitem[Bell \& de Jong(2000)]{Bell2000} Bell, E.~F. \& de Jong, R.~S.\ 2000, \mnras, 312, 497. doi:10.1046/j.1365-8711.2000.03138.x
\bibitem[Bernardi et al.(2007)]{Bernardi2007} Bernardi, M., Hyde, J.~B., Sheth, R.~K., et al.\ 2007, \aj, 133, 1741. doi:10.1086/511783
\bibitem[Bernardi et al.(2011a)]{Bernardi2011a} Bernardi, M., Roche, N., Shankar, F., et al.\ 2011a, \mnras, 412, 684. doi:10.1111/j.1365-2966.2010.17984.x
\bibitem[Bernardi et al.(2011b)]{Bernardi2011b} Bernardi, M., Roche, N., Shankar, F., et al.\ 2011b, \mnras, 412, L6. doi:10.1111/j.1745-3933.2010.00982.x
\bibitem[Blanton et al.(2011)]{Blanton2011} Blanton, M.~R., Kazin, E., Muna, D., et al.\ 2011, \aj, 142, 31. doi:10.1088/0004-6256/142/1/31
\bibitem[Boselli et al.(2008)]{Boselli2008} Boselli, A., Boissier, S., Cortese, L., et al.\ 2008, \apj, 674, 742. doi:10.1086/525513
\bibitem[Boselli \& Gavazzi(2006)]{Boselli2006} Boselli, A. \& Gavazzi, G.\ 2006, \pasp, 118, 517. doi:10.1086/500691
\bibitem[Boylan-Kolchin et al.(2009)]{Boylan2009} Boylan-Kolchin, M., Springel, V., White, S.~D.~M., et al.\ 2009, \mnras, 398, 1150. doi:10.1111/j.1365-2966.2009.15191.x
\bibitem[Cebri{\'a}n \& Trujillo(2014)]{Cebrian2014} Cebri{\'a}n, M. \& Trujillo, I.\ 2014, \mnras, 444, 682. doi:10.1093/mnras/stu1375
\bibitem[Chang et al.(2015)]{Chang2015} Chang, Y.-Y., van der Wel, A., da Cunha, E., et al.\ 2015, \apjs, 219, 8. doi:10.1088/0067-0049/219/1/8
\bibitem[Chen et al.(2022)]{Chen2022} Chen, X., Lin, Z., Kong, X., et al.\ 2022, \apj, 933, 228. doi:10.3847/1538-4357/ac75b4
\bibitem[Colbert et al.(2001)]{Colbert2001} Colbert, J.~W., Mulchaey, J.~S., \& Zabludoff, A.~I.\ 2001, \aj, 121, 808. doi:10.1086/318758
\bibitem[Cooper et al.(2012)]{Cooper2012} Cooper, M.~C., Griffith, R.~L., Newman, J.~A., et al.\ 2012, \mnras, 419, 3018. doi:10.1111/j.1365-2966.2011.19938.x
\bibitem[Covington et al.(2011)]{Covington2011} Covington, M.~D., Primack, J.~R., Porter, L.~A., et al.\ 2011, \mnras, 415, 3135. doi:10.1111/j.1365-2966.2011.18926.x
\bibitem[Dahlen et al.(2007)]{Dahlen2007} Dahlen, T., Mobasher, B., Dickinson, M., et al.\ 2007, \apj, 654, 172. doi:10.1086/508854
\bibitem[Delaye et al.(2014)]{Delaye2014} Delaye, L., Huertas-Company, M., Mei, S., et al.\ 2014, \mnras, 441, 203. doi:10.1093/mnras/stu496
\bibitem[De Rijcke et al.(2003)]{DeRijcke2003} De Rijcke, S., Dejonghe, H., Zeilinger, W.~W., et al.\ 2003, \aap, 400, 119. doi:10.1051/0004-6361:20021866
\bibitem[Desroches et al.(2007)]{Desroches2007} Desroches, L.-B., Quataert, E., Ma, C.-P., et al.\ 2007, \mnras, 377, 402. doi:10.1111/j.1365-2966.2007.11612.x
\bibitem[Ferreira et al.(2020)]{Ferreira2020} Ferreira, L., Conselice, C.~J., Duncan, K., et al.\ 2020, \apj, 895, 115. doi:10.3847/1538-4357/ab8f9b
\bibitem[Fujita(2004)]{Fujita2004} Fujita, Y.\ 2004, \pasj, 56, 29. doi:10.1093/pasj/56.1.29
\bibitem[Gadotti(2008)]{Gadotti2008} Gadotti, D.~A.\ 2008, \mnras, 384, 420. doi:10.1111/j.1365-2966.2007.12723.x
\bibitem[Graham \& Guzm{\'a}n(2003)]{GG2003} Graham, A.~W. \& Guzm{\'a}n, R.\ 2003, \aj, 125, 2936. doi:10.1086/374992
\bibitem[Graham et al.(2003)]{Graham2003} Graham, A.~W., Jerjen, H., \& Guzm{\'a}n, R.\ 2003, \aj, 126, 1787. doi:10.1086/378166
\bibitem[Gunn \& Gott(1972)]{Gunn1972} Gunn, J.~E. \& Gott, J.~R.\ 1972, \apj, 176, 1. doi:10.1086/151605
\bibitem[Guo et al.(2011)]{Guo2011} Guo, Q., White, S., Boylan-Kolchin, M., et al.\ 2011, \mnras, 413, 101. doi:10.1111/j.1365-2966.2010.18114.x
\bibitem[Haines et al.(2007)]{Haines2007} Haines, C.~P., Gargiulo, A., La Barbera, F., et al.\ 2007, \mnras, 381, 7. doi:10.1111/j.1365-2966.2007.12189.x
\bibitem[Hern{\'a}ndez-Toledo et al.(2008)]{Hernandez2008} Hern{\'a}ndez-Toledo, H.~M., V{\'a}zquez-Mata, J.~A., Mart{\'\i}nez-V{\'a}zquez, L.~A., et al.\ 2008, \aj, 136, 2115. doi:10.1088/0004-6256/136/5/2115
\bibitem[Hernquist(1989)]{Hernquist1989} Hernquist, L.\ 1989, \nat, 340, 687. doi:10.1038/340687a0
\bibitem[Holden et al.(2012)]{Holden2012} Holden, B.~P., van der Wel, A., Rix, H.-W., et al.\ 2012, \apj, 749, 96. doi:10.1088/0004-637X/749/2/96
\bibitem[Hopkins et al.(2008)]{Hopkins2008} Hopkins, P.~F., Hernquist, L., Cox, T.~J., et al.\ 2008, \apjs, 175, 356. doi:10.1086/524362
\bibitem[Huertas-Company et al.(2013)]{Huertas-Company2013} Huertas-Company, M., Mei, S., Shankar, F., et al.\ 2013, \mnras, 428, 1715. doi:10.1093/mnras/sts150
\bibitem[Hyde \& Bernardi(2009)]{Hyde2009} Hyde, J.~B. \& Bernardi, M.\ 2009, \mnras, 394, 1978. doi:10.1111/j.1365-2966.2009.14445.x
\bibitem[Kawinwanichakij et al.(2017)]{Kawinwanichakij2017} Kawinwanichakij, L., Papovich, C., Quadri, R.~F., et al.\ 2017, \apj, 847, 134. doi:10.3847/1538-4357/aa8b75
\bibitem[Kawinwanichakij et al.(2021)]{Kawinwanichakij2021} Kawinwanichakij, L., Silverman, J.~D., Ding, X., et al.\ 2021, \apj, 921, 38. doi:10.3847/1538-4357/ac1f21
\bibitem[Kelkar et al.(2015)]{Kelkar2015} Kelkar, K., Arag{\'o}n-Salamanca, A., Gray, M.~E., et al.\ 2015, \mnras, 450, 1246. doi:10.1093/mnras/stv670
\bibitem[Ko et al.(2013)]{Ko2013} Ko, J., Hwang, H.~S., Lee, J.~C., et al.\ 2013, \apj, 767, 90. doi:10.1088/0004-637X/767/1/90
\bibitem[Kuchner et al.(2017)]{Kuchner2017} Kuchner, U., Ziegler, B., Verdugo, M., et al.\ 2017, \aap, 604, A54. doi:10.1051/0004-6361/201630252
\bibitem[Kuntschner et al.(2002)]{Kuntschner2002} Kuntschner, H., Smith, R.~J., Colless, M., et al.\ 2002, \mnras, 337, 172. doi:10.1046/j.1365-8711.2002.05897.x
\bibitem[Lacerna et al.(2016)]{Lacerna2016} Lacerna, I., Hern{\'a}ndez-Toledo, H.~M., Avila-Reese, V., et al.\ 2016, \aap, 588, A79. doi:10.1051/0004-6361/201527844
\bibitem[Lange et al.(2015)]{Lange2015} Lange, R., Driver, S.~P., Robotham, A.~S.~G., et al.\ 2015, \mnras, 447, 2603. doi:10.1093/mnras/stu2467
\bibitem[Lani et al.(2013)]{Lani2013} Lani, C., Almaini, O., Hartley, W.~G., et al.\ 2013, \mnras, 435, 207. doi:10.1093/mnras/stt1275
\bibitem[Larson et al.(1980)]{Larson1980} Larson, R.~B., Tinsley, B.~M., \& Caldwell, C.~N.\ 1980, \apj, 237, 692. doi:10.1086/157917
\bibitem[Lee et al.(2015)]{Lee2015} Lee, S.-K., Im, M., Kim, J.-W., et al.\ 2015, \apj, 810, 90. doi:10.1088/0004-637X/810/2/90
\bibitem[Lisker et al.(2006)]{Lisker2006} Lisker, T., Grebel, E.~K., \& Binggeli, B.\ 2006, \aj, 132, 497. doi:10.1086/505045
\bibitem[Lopes et al.(2016)]{Lopes2016} Lopes, P.~A.~A., Rembold, S.~B., Ribeiro, A.~L.~B., et al.\ 2016, \mnras, 461, 2559. doi:10.1093/mnras/stw1497
\bibitem[Madau \& Dickinson(2014)]{Madau2014} Madau, P. \& Dickinson, M.\ 2014, \araa, 52, 415. doi:10.1146/annurev-astro-081811-125615
\bibitem[Masters et al.(2012)]{Masters2012} Masters, K.~L., Nichol, R.~C., Haynes, M.~P., et al.\ 2012, \mnras, 424, 2180. doi:10.1111/j.1365-2966.2012.21377.x
\bibitem[McIntosh et al.(2014)]{McIntosh2014} McIntosh, D.~H., Wagner, C., Cooper, A., et al.\ 2014, \mnras, 442, 533. doi:10.1093/mnras/stu808
\bibitem[Meert et al.(2015)]{Meert2015} Meert, A., Vikram, V., \& Bernardi, M.\ 2015, \mnras, 446, 3943. doi:10.1093/mnras/stu2333
\bibitem[Moore et al.(1996)]{Moore1996} Moore, B., Katz, N., Lake, G., et al.\ 1996, \nat, 379, 613. doi:10.1038/379613a0
\bibitem[Morishita et al.(2017)]{Morishita2017} Morishita, T., Abramson, L.~E., Treu, T., et al.\ 2017, \apj, 835, 254. doi:10.3847/1538-4357/835/2/254
\bibitem[Morishita et al.(2014)]{Morishita2014} Morishita, T., Ichikawa, T., \& Kajisawa, M.\ 2014, \apj, 785, 18. doi:10.1088/0004-637X/785/1/18
\bibitem[Mosleh et al.(2013)]{Mosleh2013} Mosleh, M., Williams, R.~J., \& Franx, M.\ 2013, \apj, 777, 117. doi:10.1088/0004-637X/777/2/117
\bibitem[Mosleh et al.(2011)]{Mosleh2011} Mosleh, M., Williams, R.~J., Franx, M., et al.\ 2011, \apj, 727, 5. doi:10.1088/0004-637X/727/1/5
\bibitem[Moutard et al.(2018)]{Moutard2018} Moutard, T., Sawicki, M., Arnouts, S., et al.\ 2018, \mnras, 479, 2147. doi:10.1093/mnras/sty1543
\bibitem[Mowla et al.(2019)]{Mowla2019} Mowla, L., van der Wel, A., van Dokkum, P., et al.\ 2019, \apjl, 872, L13. doi:10.3847/2041-8213/ab0379
\bibitem[Nedkova et al.(2021)]{Nedkova2021} Nedkova, K.~V., H{\"a}u{\ss}ler, B., Marchesini, D., et al.\ 2021, \mnras, 506, 928. doi:10.1093/mnras/stab1744
\bibitem[Niemi et al.(2010)]{Niemi2010} Niemi, S.-M., Hein{\"a}m{\"a}ki, P., Nurmi, P., et al.\ 2010, \mnras, 405, 477. doi:10.1111/j.1365-2966.2010.16457.x
\bibitem[Oogi \& Habe(2013)]{Oogi2013} Oogi, T. \& Habe, A.\ 2013, \mnras, 428, 641. doi:10.1093/mnras/sts047
\bibitem[Paulino-Afonso et al.(2017)]{Paulino2017} Paulino-Afonso, A., Sobral, D., Buitrago, F., et al.\ 2017, \mnras, 465, 2717. doi:10.1093/mnras/stw2933
\bibitem[Pedraz et al.(2002)]{Pedraz2002} Pedraz, S., Gorgas, J., Cardiel, N., et al.\ 2002, \mnras, 332, L59. doi:10.1046/j.1365-8711.2002.05565.x
\bibitem[Peng et al.(2010)]{Peng2010} Peng, Y.-. jie ., Lilly, S.~J., Kova{\v{c}}, K., et al.\ 2010, \apj, 721, 193. doi:10.1088/0004-637X/721/1/193
\bibitem[Reda et al.(2004)]{Reda2004} Reda, F.~M., Forbes, D.~A., Beasley, M.~A., et al.\ 2004, \mnras, 354, 851. doi:10.1111/j.1365-2966.2004.08250.x
\bibitem[Reda et al.(2005)]{Reda2005} Reda, F.~M., Forbes, D.~A., \& Hau, G.~K.~T.\ 2005, \mnras, 360, 693. doi:10.1111/j.1365-2966.2005.09058.x
\bibitem[Reduzzi et al.(1996)]{Reduzzi1996} Reduzzi, L., Longhetti, M., \& Rampazzo, R.\ 1996, \mnras, 282, 149. doi:10.1093/mnras/282.1.149
\bibitem[Ribeiro et al.(2016)]{Ribeiro2016} Ribeiro, B., Le F{\`e}vre, O., Tasca, L.~A.~M., et al.\ 2016, \aap, 593, A22. doi:10.1051/0004-6361/201628249
\bibitem[Robaina et al.(2012)]{Robaina2012} Robaina, A.~R., Hoyle, B., Gallazzi, A., et al.\ 2012, \mnras, 427, 3006. doi:10.1111/j.1365-2966.2012.21804.x
\bibitem[Robertson et al.(2006)]{Robertson2006} Robertson, B., Cox, T.~J., Hernquist, L., et al.\ 2006, \apj, 641, 21. doi:10.1086/500360
\bibitem[Rodriguez-Gomez et al.(2015)]{Rodriguez2015} Rodriguez-Gomez, V., Genel, S., Vogelsberger, M., et al.\ 2015, \mnras, 449, 49. doi:10.1093/mnras/stv264
\bibitem[Shankar et al.(2013)]{Shankar2013} Shankar, F., Marulli, F., Bernardi, M., et al.\ 2013, \mnras, 428, 109. doi:10.1093/mnras/sts001
\bibitem[Shankar et al.(2014)]{Shankar2014} Shankar, F., Mei, S., Huertas-Company, M., et al.\ 2014, \mnras, 439, 3189. doi:10.1093/mnras/stt2470
\bibitem[Shen et al.(2003)]{Shen2003} Shen, S., Mo, H.~J., White, S.~D.~M., et al.\ 2003, \mnras, 343, 978. doi:10.1046/j.1365-8711.2003.06740.x
\bibitem[Strauss et al.(2002)]{Strauss2002} Strauss, M.~A., Weinberg, D.~H., Lupton, R.~H., et al.\ 2002, \aj, 124, 1810. doi:10.1086/342343
\bibitem[Toloba et al.(2011)]{Toloba2011} Toloba, E., Boselli, A., Cenarro, A.~J., et al.\ 2011, \aap, 526, A114. doi:10.1051/0004-6361/201015344
\bibitem[van Dokkum et al.(2015)]{vanDokkum2015} van Dokkum, P.~G., Nelson, E.~J., Franx, M., et al.\ 2015, \apj, 813, 23. doi:10.1088/0004-637X/813/1/23
\bibitem[Wake et al.(2017)]{Wake2017} Wake, D.~A., Bundy, K., Diamond-Stanic, A.~M., et al.\ 2017, \aj, 154, 86. doi:10.3847/1538-3881/aa7ecc
\bibitem[Wetzel et al.(2013)]{Wetzel2013} Wetzel, A.~R., Tinker, J.~L., Conroy, C., et al.\ 2013, \mnras, 432, 336. doi:10.1093/mnras/stt469
\bibitem[Williams et al.(2010)]{Williams2010} Williams, R.~J., Quadri, R.~F., Franx, M., et al.\ 2010, \apj, 713, 738. doi:10.1088/0004-637X/713/2/738
\bibitem[Yoon et al.(2017)]{Yoon2017} Yoon, Y., Im, M., \& Kim, J.-W.\ 2017, \apj, 834, 73. doi:10.3847/1538-4357/834/1/73
\end{thebibliography}
\end{document}